\documentclass[prd,twocolumn,floatfix,amsmath,nofootinbib,amssymb,floatfix,preprintnumbers,showkeys]{revtex4}
\usepackage{graphicx,color,dcolumn,booktabs,bm,multirow}
\usepackage{longtable,lscape}
\usepackage{txfonts}
\usepackage{overpic}
\usepackage{amssymb}
\usepackage{array}
\usepackage{indentfirst}
\usepackage{feynmf}   %{feynmp}
\usepackage{slashed}  %for Feynman symbols
\usepackage{cases}
\usepackage{color}
\usepackage{multirow}
\usepackage{epstopdf}
\usepackage{tabularx}
\usepackage{graphicx,color,dcolumn,booktabs,bm}
\usepackage[colorlinks,
citecolor=blue,
anchorcolor=red,
menucolor=red,
linkcolor=red,
filecolor=red,
runcolor=red,
urlcolor=blue,
frenchlinks=red]{hyperref}
\usepackage{braket}
\usepackage{CJKutf8}

\begin{document}
\begin{CJK}{UTF8}{gbsn}

\title{Dipion transitions from $X(3872)$ to $\chi_{cJ}\ (J=0,1,2)$ }
\author{Qi Wu$^{1}$}%\email{wuqi@htu.edu.cn}
\author{Zhong-Quan Sun$^{1}$}%\email{2638949959@qq.com}
\author{Dian-Yong Chen$^{2,3}$}\email{chendy@seu.edu.cn}
\author{Shi-Dong Liu$^{4}$} \email{liusd@qfnu.edu.cn}
\author{Gang Li$^{4}$ }\email{gli@qfnu.edu.cn}
\affiliation{$^1$ School of Physics, Henan Normal University, Xinxiang 453007, China}
\affiliation{$^2$ School of Physics, Southeast University, Nanjing 210094, China}
\affiliation{$^3$ Lanzhou Center for Theoretical Physics, Lanzhou University, Lanzhou 730000, China}
\affiliation{$^4$ College of Physics and Engineering, Qufu Normal University, Qufu 273165, China}
\date{\today}

\begin{abstract}
In this work, we investigate the dipion transition processes $X(3872)\to \pi \pi \chi_{cJ} (J=0,1,2)$ within the framework of heavy hadron chiral perturbation theory, treating $X(3872)$ as a molecular state composed of $D\bar{D}^*$+ H.c. components. By analyzing the box and triangle loop diagrams with the nonrelativistic effective field theory power-counting rule, we demonstrate that box diagrams dominate these dipion transition processes. Branching ratios are calculated as functions of the mixing angle $\theta$, which parametrizes the neutral and charged meson compositions of the $X(3872)$. Our results indicate that the branching fractions for $X(3872)\to\pi\pi\chi_{c0}$, $X(3872)\to \pi\pi\chi_{c1}$, and $X(3872)\to \pi\pi\chi_{c2}$ are of the
orders of $10^{-4}$, $10^{-3}$, and $10^{-5}$, respectively. We also predict the ratios ${\mathcal{B}[X(3872)\rightarrow \pi\pi\chi_{c0/2}]}/{\mathcal{B}[X(3872)\rightarrow \pi\pi\chi_{c1}]}$ and ${\mathcal{B}[X(3872)\rightarrow \pi^+\pi^-\chi_{cJ}]}/{\mathcal{B}[X(3872)\rightarrow \pi^0\pi^0\chi_{cJ}]}$. The latter deviates from isospin-symmetry expectations, revealing various degrees of isospin violation. By studying the $\pi^+\pi^-$ and $\pi^+\chi_{cJ}$ invariant mass spectra, we find a double-bump structure in the $\pi^
+ \pi^-$ invariant mass distributions of the process $X(3872)\to \pi^+\pi^-\chi_{c1}$ and $\pi^+\chi_{c0}$ invariant mass distribution of the process $X(3872)\to \pi^+\pi^-\chi_{c0}$, which could be tested by future experimental measurements.
\end{abstract}
\maketitle

\section{introduction}
\label{Sec:introduction}
The exotic hadron state $X(3872)$, discovered by the Belle Collaboration in 2003 in the decay $B^\pm\to K^\pm X(3872)$ with $X(3872)\to \pi^+ \pi^- J/\psi$ ~\cite{Belle:2003nnu}, remains one of the most intriguing puzzles in quantum chromodynamics. It challenges conventional quark model spectroscopy and marks a new era of hadron spectroscopy (see Refs.~\cite{Chen:2016qju,Hosaka:2016pey,Lebed:2016hpi,Esposito:2016noz,Guo:2017jvc,Ali:2017jda,Olsen:2017bmm,Karliner:2017qhf,Yuan:2018inv,Dong:2017gaw,Liu:2019zoy,
Liu:2024uxn} for recent reviews). The $J^{PC}$ quantum numbers of $X(3872)$ have been determined to be $1^{++}$~\cite{LHCb:2015jfc}, which is allowed for the \textit{P}-wave conventional charmonia. However, its mass, measured to be $(3871.69 \pm 0.17)$ MeV, is approximately 80 MeV lower than the prediction from Godfrey-Isgur quark model~\cite{Godfrey:1985xj}, rendering it too light to be the $\chi_{c1}(2P)$ state.
Another novel nature of $X(3872)$ is its mass being extremely close to the $D^0\bar{D}^{*0}$ threshold. This proximity strongly suggests that this state possesses a significant $D^0\bar{D}^{*0}$ molecular component~\cite{Tornqvist:1993ng,Voloshin:2003nt,Swanson:2003tb,Tornqvist:2004qy,Fleming:2007rp,Liu:2008fh}, providing a natural explanation for its exotic properties.

Within the molecular scheme, both the decay and production behaviors of the $X(3872)$ can be naturally accounted for~\cite{Swanson:2004pp,Braaten:2003he,Braaten:2004ai,Braaten:2005ai,Liu:2006df,Dubynskiy:2007tj,Fleming:2008yn,Dong:2008gb,Bignamini:2009sk,Fleming:2011xa,
Aceti:2012cb,Margaryan:2013tta,Guo:2013zbw,Guo:2014hqa,Zhou:2019swr,Wu:2021udi,Meng:2021kmi,Wang:2022qxe,Wu:2023rrp}. The strongest evidence supporting the molecular hypothesis is the nearly equal branching fractions of the $X(3872)\to \pi^+\pi^- J/\psi$ and $X(3872)\to \pi^+\pi^-\pi^0 J/\psi$~\cite{Belle:2005lfc,BESIII:2019qvy,BaBar:2010wfc}:
\begin{eqnarray}
\renewcommand\arraystretch{1.35}
\frac{\mathcal{B}[X\to \pi^+ \pi^- \pi^0 J/\psi]}{\mathcal{B}[X\to \pi^+ \pi^- J/\psi]}=
\left\{
\begin{array}{ll}
	1.0\pm 0.4 \pm 0.3  & \mathrm{Belle}\\
	1.43^{+0.28}_{-0.23} & \mathrm{BESIII} \\
	0.8\pm0.3 & \mathrm{BABAR}\\
\end{array}
\right.
\label{Eq:Ratio1}
\end{eqnarray}
Measurements of the multipion invariant mass distributions reveal the dominate decay pathways: The $\pi^+ \pi^-J/\psi$ final state proceeds predominantly via $\rho^0 J/\psi$~\footnote{Actually, the BESIII Collaboration has measured the $\rho^0$ and $\omega$ contributions to the $X(3872)\to \pi^+ \pi^- J/\psi$ decays, and the $\rho^0$ contribution accounts for $78.6\%$ of the total rate~\cite{LHCb:2022jez}.}, while the $\pi^+ \pi^- \pi^0 J/\psi$ final state occurs primarily through $J/\psi\omega$. This suggests that the $X(3872)$ couples to both isospin $I=0$ and $I=1$ channels with comparable strength, resulting in large isospin violation in these decay patterns. It is reasonable, since the charged threshold $D^+ D^{*-}$+ c.c. is about 8 MeV above the neutral threshold, and the mass of $X(3872)$ lies so close to the neutral threshold. Numerous studies suggest that the weight of the $D^0 \bar{D}^{*0}$ component in $X(3872)$ is over $80\%$~\cite{Yamaguchi:2019vea,Wu:2021udi,Meng:2021kmi,Wu:2023rrp,Wang:2023ovj,Song:2023pdq}, despite the important role of the charged $D\bar{D}^{*}$ components in describing the ratio of the branching fractions of $\rho J/\psi$ and $\omega J/\psi$ channels~\cite{Gamermann:2009uq,Gamermann:2007fi,Gamermann:2007mu,Gamermann:2009fv,Gamermann:2010ga}.

In addition to the $X(3872)\to \rho^0 J/\psi$, the decays $X(3872)\to \pi^0 \chi_{cJ}$ with $J=0,1,2$ are isospin violated and presumed highly suppressed. It is predicted that the decay rates of pionic transitions from $X(3872)$ to $\chi_{cJ}$ are sensitive to the inner structure of the $X(3872)$~\cite{Dubynskiy:2007tj,Fleming:2008yn,Fleming:2011xa}.  For example, the authors of Ref.~\cite{Dubynskiy:2007tj} first proposed the pionic transitions between $X(3872)$ and $\chi_{cJ}$ and investigated their relative rates under different interpretations of $X(3872)$, particularly, the pure charmonium and four-quark and molecular configurations. Their analysis concluded that (i) the decay rates are highly suppressed (especially for single-pion transitions); (ii) $X(3872)\to \pi\pi \chi_{c1}$ dominates for the charmonium scheme; and (iii) single-pion transitions are significantly enhanced in four-quark and molecular scenarios. The authors in Ref.~\cite{Fleming:2008yn} investigated the decays of the $X(3872)$ to \textit{P}-wave charmonia under the molecular picture and revisited the decay rate of $X(3872)\to \pi\pi \chi_{c1}$ in Ref.~\cite{Fleming:2011xa} by considering a low-energy effective theory for the $X(3872)$ (XEFT). The former work predicted that dipionic transitions are suppressed by 6 orders of magnitude relative to single-pion transitions for $J=0,2$, with $\mathcal{B}[X(3872)\to \pi^+ \pi^- \chi_{cJ}]/\mathcal{B}[X(3872)\to \pi^0 \pi^0 \chi_{cJ}] \simeq 2$. The latter work revisited the ratio to be $\mathcal{B}[X(3872)\to \pi^0 \pi^0 \chi_{c1}]/\mathcal{B}[X(3872)\to \pi^0 \chi_{c1}]=10^{-3}$. Considering the $X(3872)$ as a superposition dominated by the molecular $D^0 D^{*0}$ component with additional contributions from $D^\pm D^{*\mp}$, $\omega J/\psi$, and $\rho J/\psi$ states, the authors in Ref.~\cite{Dong:2009yp} calculated the decays $X(3872)\to \pi \chi_{cJ}$ and $X(3872)\to \pi\pi \chi_{cJ}$ as well as the pionic transitions from $X(3872)$ to $J/\psi$ by using the estimated couplings $g_{X\psi\omega}$ and $g_{X\psi\rho}$. The full structure-dependent decay pattern of the $X(3872)$ established in Ref.~\cite{Dong:2009yp} may help determine its hadronic composition in current and future experiments.

In 2019, the BESIII Collaboration reported the first observation of $X(3872)\to \pi^0 \chi_{c1}$, and the central value of the ratio $\mathcal{B}[X(3872)\to \pi^0 \chi_{c1}]/\mathcal{B}[X(3872)\to \pi^+\pi^- J/\psi]$ was measured to be $0.88$~\cite{BESIII:2019esk}. Using the Particle Data Group value $\mathcal{B}[X(3872)\to \pi^+\pi^- J/\psi]=4.3\%$~\cite{ParticleDataGroup:2024cfk}, it implies $\mathcal{B}[X(3872)\to \pi^0 \chi_{c1}] =3.8\%$. This branching ratio is comparable to those for $X(3872)\to \pi\pi(\pi)J/\psi$. However, other pionic transitions from $X(3872)$ to \textit{P}-wave charmonium are kept unknown, and only some upper values have been measured. In 2022, the BESIII Collaboration searched for $X(3872)\to \pi^0 \chi_{c0}$ and $X(3872)\to \pi^0 \pi^0 \chi_{c0}$ via $e^+ e^- \to \gamma X(3872)$~\cite{BESIII:2022kow}.  The upper limits of these ratios were measured to be
\begin{eqnarray}
\renewcommand\arraystretch{1.35}
\frac{\mathcal{B}[X\to  [\pi] \chi_{c0}]}{\mathcal{B}[X\to \pi^+ \pi^- J/\psi]}<
\left\{
\begin{array}{ll}
	3.6  & [\pi]=\pi^0\\
	0.56 & [\pi]=\pi^+\pi^- \\
	1.7  & [\pi]=\pi^0\pi^0\\
\end{array}
\right.
\label{Eq:Ratio2}
\end{eqnarray}
at $90\%$ confidence level. Recently, the BESIII Collaboration made further progress in the study of $X(3872)\to \pi\pi \chi_{c1,2}$, and the upper limits of the ratios were measured to be~\cite{BESIII:2023eeb,BESIII:2024ilt}
\begin{eqnarray}
\renewcommand\arraystretch{1.35}
\frac{\mathcal{B}[X\to  [\pi\pi] \chi_{cJ}]}{\mathcal{B}[X\to \pi^+ \pi^- J/\psi]}<
\left\{
\begin{array}{ll}
	0.18  & [\pi\pi]=\pi^+\pi^-,J=1\\
	1.1   & [\pi\pi]=\pi^0\pi^0,J=1 \\
	0.5   & [\pi\pi]=\pi^0\pi^0,J=2\\
\end{array}
\right.
\label{Eq:Ratio3}
\end{eqnarray}
at $90\%$ confidence level.

The BESIII measurement inspired theoretical studies on the dipionic transitions from $X(3872)$ to $\chi_{cJ}$. In Ref.~\cite{Achasov:2024anu}, the decays $X(3872)\to \pi^0 \pi^0 \chi_{c1}$ and $X(3872)\to \pi^+ \pi^- \chi_{c1}$ were estimated in the triangle loop model by considering the $X(3872)$ meson as a $\chi_{c1}(2P)$ charmonium state. The obtained branching fractions of the $X(3872)\to \pi^+ \pi^- \chi_{c1}$ and $X(3872)\to \pi^0 \pi^0 \chi_{c1}$ decays are in order $10^{-5}-10^{-4}$, and their ratio is estimated to be 1.1. However, the contribution from the triangle loop diagrams to the dipionic transition processes of the $X(3872)$ and its heavy quark flavor and spin symmetry partners is expected to be suppressed relative to the box diagrams~\cite{Jia:2023pud,Liu:2024ogo,Cai:2025inq}. In Ref.~\cite{Jia:2023pud}, based on a power-counting scheme in heavy hadron chiral perturbation theory (HH$\chi$PT), the contributions of the triangle diagrams to the decay widths were estimated to be 1¨C3 magnitudes smaller than the box diagram contributions for $X_b \to \pi\pi\chi_{bJ}$. Similar conclusions were also drawn in the study of dipionic transitions of $X_2$ in Ref.~\cite{Liu:2024ogo} by using a power-counting rule, where the ratio of triangle diagrams and box diagrams is of the order of $\mathcal{O}(0.1)$, although the power counting somewhat overestimates the suppression on the triangle loops for the $X_2\to \pi\pi \chi_{cJ}$. The dipionic transitions from $X(3872)$ to $\eta_c$ were studied in Ref.~\cite{Cai:2025inq} using an effective Lagrangian approach, and the numerical results reveal that the decay width from the triangle diagrams is one order of magnitude smaller than that from box diagrams.

In Ref.~\cite{Wu:2021udi}, we investigated the isospin breaking in the hidden charm decay processes, $X(3872)\to \rho(\omega) J/\psi$ and $X(3872)\to \pi^0 \chi_{cJ}$, with the triangle loop mechanism, and we found that isospin-symmetry breaking stems from two aspects, which are the mass difference between charged and neutral charm mesons in the meson loop, and the different fraction of the charged and neutral components in the molecular states. In the present work, we study dipion transition processes $X(3872) \to \pi\pi \chi_{cJ}(J=0,1,2)$ in HH$\chi$PT by considering box intermediate meson-loop contributions. This allows us to make model-independent predictions and provide quantitative estimates of different contributions based on the power counting. Under the molecular picture of $X(3872)$, we calculate the relative decay rates to different final states. These ratios serve as critical tests of the molecular interpretation, as heavy quark symmetry constrains their expected values.

The paper is organized as follows. In Sec.~\ref{sec:formula}, we present the details of the theoretical framework adopted in the present estimations. Numerical results and the relevant discussions are given in Sec.~\ref{sec:results}, with conclusions summarized in Sec.~\ref{sec:summary}.

\section{Theoretical Framework}
\label{sec:formula}

\subsection{Effective Lagrangians}

\begin{figure}
    \centering    \includegraphics[width=0.98\linewidth]{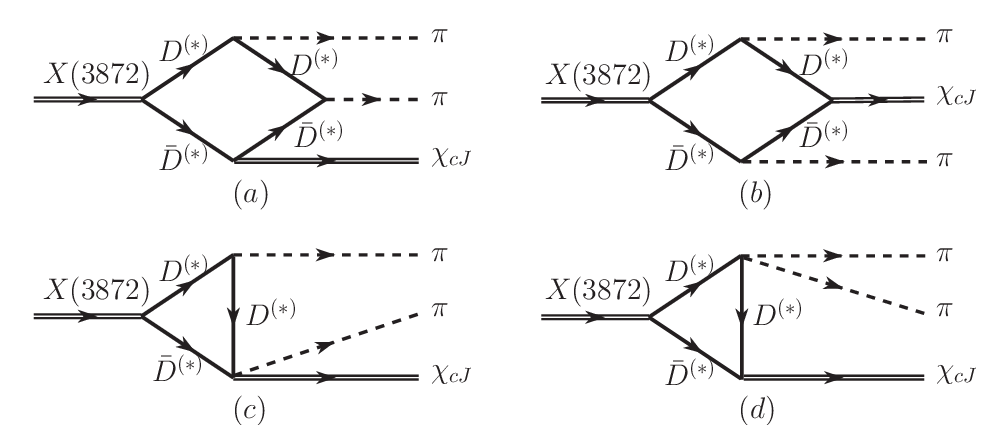}
    \caption{Box [(a), (b)] and triangle [(c), (d)] diagrams contributing to the dipionic transitions from $X(3872)$ to $\chi_{cJ}$.}
    \label{fig:feyndiags}
\end{figure}

The sketch diagrams contributing to the dipion decay processes $X(3872) \to \pi \pi \chi_{cJ}\ (J=0,1,2)$ are presented in Fig.~\ref{fig:feyndiags}, where diagrams (a),(b) and (c),(d) are box and triangle diagrams, respectively. Following Ref.~\cite{Jia:2023pud}, we estimate the relative contributions of box and triangle diagrams using a NREFT power-counting rule~\cite{Guo:2009wr,Guo:2010ak}. In this framework, the momentum, nonrelativistic energy, integral measure, and heavy meson propagator count as $v$, $v^2$, $v^5$, and $1/v^2$, respectively. The \textit{S}-wave vertices are velocity independent, while $P$-wave vertices scale as either $v$ or external momentum. As for Figs.~\ref{fig:feyndiags}(a) and \ref{fig:feyndiags}(b), the vertices $XD\bar{D}^*/XD^* \bar{D}$ and $\chi_{cJ}D^{(*)}\bar{D}^{(*)}$ are in \textit{S}-wave (scaling $\sim 1$). The $D^{*}D^{(*)}\pi$ and $\chi_{cJ}D^{(*)}\bar{D}^{(*)}\pi$ vertices have \textit{P}-wave couplings, introducing a factor $\vec{q}$ to the amplitudes, where $\vec{q}$ is the pion momentum. Besides, the four interaction vertices $D^{(*)}D^{(*)}\pi\pi$ are proportional to the square of the energy of the pion. As a result, the amplitude corresponding to Figs.~\ref{fig:feyndiags}(a) and \ref{fig:feyndiags}(b) scales as $v^5 \vec{q}^2/(v^2)^4=\vec{q}^2/v^3$, and the amplitude related to Figs.~\ref{fig:feyndiags}(c) and \ref{fig:feyndiags}(d) scales as $v^5 \vec{q}^2/(v^2)^3=\vec{q}^2/v$. The ratio of the amplitude from box and triangle diagrams can be roughly estimated to be $1/v^2$, where $v\ll 1$ since it is the typical velocity of the nonrelativistic intermediate mesons. Consequently, the contribution from box diagrams to the dipionic transitions from $X(3872)$ to $\chi_{cJ}$ is much larger than that from triangle diagrams.

The power-counting analysis indicates that box diagrams dominate the dipionic transitions $X(3872)\to \pi\pi \chi_{cJ}$. We notice that the authors in Ref.~\cite{Dong:2009yp} have studied the decays $X(3872)\to \pi\pi \chi_{cJ}$ by calculating the box diagrams via a phenomenological Lagrangian approach. However, they consider only the coupling of the $X(3872)$ with a neutral component. In the following, we present the relevant effective Lagrangians associated with the box diagram calculations. The $X(3872)$ is assumed as an \textit{S}-wave molecular state with the quantum numbers of $J^{PC}=1^{++}$. As a pure hadronic molecule, the wave function of the $X(3872)$ can be written as
\begin{align}\label{eq:WFX}
        |X(3872)\rangle&=\frac{\cos{\theta}}{\sqrt{2}}|D^{\ast 0}\bar D^{0}+D^{0}\bar D^{\ast 0}\rangle \nonumber\\
        &+\frac{\sin{\theta}}{\sqrt{2}}|D^{\ast +}D^{-}+D^{+}D^{\ast -}\rangle,
    \end{align}
where $\theta$ is the mixing angle describing the proportion of neutral and charged constituents.

The coupling of the $X(3872)$ to a pair of charmed and anticharmed mesons is described by effective Lagrangian
\begin{align}
    \mathcal{L}_X&=\frac{g_n}{\sqrt{2}}X_{i}^{\dagger}(D^{\ast 0i}\bar D^{0}+D^{0}\bar D^{\ast0i})\nonumber\\
    &+\frac{g_c}{\sqrt{2}}X_{i}^{\dagger}(D^{\ast+i}D^{-}+D^{+}D^{\ast-i}), \label{Eq:LX}
\end{align}
where $g_n=|g^n_{\mathrm{NR}}|\cos{\theta}$ and $g_c=|g^c_{\mathrm{NR}}|\sin{\theta}$ are the coupling constants of the $X(3872)$ with its neutral and charged components, respectively.

The pionic couplings to heavy mesons are constrained by chiral symmetry. For the $S$-wave heavy mesons, the leading-order Lagrangian in heavy meson chiral perturbation theory is~\cite{Wise:1992hn,Hu:2005gf}
\begin{eqnarray}
\mathcal{L}_\pi&=&-\frac{g}{2}\ \mathrm{Tr}[H^\dag_{a} H_{b}\vec{\sigma}\cdot\vec{u}_{ba}]\nonumber\\
&=&\frac{\sqrt{2}g}{F}(i\epsilon^{ijk}V^{i\dag}_a V^j_b\partial^k\phi_{ba}+V^{i\dag}_aP_b\partial^i\phi_{ba}+P^\dag_a V^i_b\partial^i\phi_{ba}), \nonumber\\ \label{Eq:Lpi}
\end{eqnarray}
where the axial current is $\vec{u}=-\sqrt{2}\vec{\partial}\phi/F_\pi + \mathcal{O}(\phi^3)$. Charmed mesons with $s^P_l=1/2^+$ form a spin multiplet, which is written as $H_{a}=\vec{V}_a\cdot\vec{\sigma}+P_a$ in two-component notation~\cite{Hu:2005gf}. Here, $F_\pi$ is the pion decay constant in the chiral limit, and
\[ \phi = \begin{pmatrix}
\pi^0/\sqrt{2} & \pi^+ \\
\pi^- & -\pi^0/\sqrt{2}
\end{pmatrix}
\]
collects the pion fields.

The coupling of the \textit{P}-wave charmonia $\chi_{cJ}$ to a pair of charmed and anticharmed mesons is described as~\cite{Fleming:2008yn,Mehen:2015efa}
\begin{equation}
\begin{aligned}\mathcal{L}_{\chi}&=i\frac{g_\chi}{2} \mathrm{Tr}[\chi^{\dagger i}H_a\sigma^i\bar{H}_a]+\mathrm{H.c.}\\
&=i2g_\chi \chi_{c2}^{\dagger ij}V_a^i\bar{V}_a^j+\sqrt{2}g_\chi \chi_{c1}^{\dagger i}(V_a^i\bar{P}_a+P_a\bar{V}_a^i)\\&+\frac{i}{\sqrt{3}}g_\chi \chi_{c0}^\dagger(\vec{V}_a \cdot \vec{\bar{V}}_a+3P_a\bar{P}_a)+\mathrm{H.c.} \, ,\label{Eq:Lchi}
\end{aligned}
\end{equation}
where $\chi$ denotes the field for the \textit{P}-wave charmonia,
\begin{equation}
\chi^{i}=\sigma^{j}\left(-\chi_{c2}^{ij}-\frac{1}{\sqrt{2}}\epsilon^{ijk}\chi_{c1}^{k}+\frac{1}{\sqrt{3}}\delta^{ij}\chi_{c0}\right)+h_{c}^{i}.
\end{equation}

\subsection{Decay amplitudes}

The relevant kinematics we used for calculations are explicitly indicated in Fig. \ref{fig:feyndiagsx}.
With the Lagrangians in Eqs.~(\ref{Eq:LX})-(\ref{Eq:Lchi}), the decay amplitudes for the $X(3872)\to \pi^0 \pi^0 \chi_{cJ}$ read as
\begin{eqnarray}
&&\mathcal{M}_{\pi^0\pi^0\chi_{c0}}=\sqrt{\frac23}\frac1{F_\pi^2} g^2 g_1 q_1^i q_2^j \varepsilon^k(p) \epsilon_{i j k}\nonumber\\
&&\times\Big[g_n\Big(I_{1}(D^0,\bar{D}^{*0},\bar{D}^{*0},D^{*0})+3I_{1}(D^{*0},\bar{D}^{0},\bar{D}^{0},D^{*0}) \nonumber\\
&&-3I_{1}^{\prime}(D^{*0},\bar{D}^{0},\bar{D}^{0},D^{*0})-I_1^\prime(D^0,\bar{D}^{*0},\bar{D}^{*0},D^{*0})\nonumber\\
&&+I_2(D^{*0},\bar{D}^0,\bar{D}^{*0},D^{*0})-I_{2}(D^0,\bar{D}^{*0},\bar{D}^{*0},D^{*0})\Big)\nonumber\\
&&+g_c\Big(I_{1}(D^{+},D^{*-},D^{*-},D^{*+})+3I_{1}(D^{*+},D^-,D^-,D^{*+}) \nonumber\\
&&-3I_{1}^{\prime}(D^{*+},D^-,D^-,D^{*+})-I_1^\prime(D^{+},D^{*-},D^{*-},D^{*+})\nonumber\\
&&+I_2(D^{*+},D^-,D^{*-},D^{*+})-I_{2}(D^{+},D^{*-},D^{*-},D^{*+})\Big)\Big], \label{Eq:amp1}
\end{eqnarray}
\begin{eqnarray}
&&\mathcal{M}_{\pi^0\pi^0\chi_{c1}}=\frac{2}{F_\pi^2} g^2 g_1 \Big\{\vec{q}_2\cdot \vec{\varepsilon}(p) \vec{q}_1\cdot \vec{\varepsilon}(q_3)\times\nonumber\\
&&\Big[g_n\Big(I_1(D^{*0},\bar{D}^{0},D^{*0},D^{*0})-I^\prime_1(D^{*0},\bar{D}^{0},\bar{D}^{*0},D^{0})\nonumber\\
&&-I_2(D^{0},\bar{D}^{*0},D^{0},D^{*0})\Big)+g_c\Big(I_{1}(D^{*+},D^-,D^{*+},D^{*+})\nonumber\\
&&-I^\prime_1(D^{*+},D^-,D^{*-},D^{+})-I_2(D^{+},D^{*-},D^-,D^{*+})\Big)\Big]\nonumber\\
&&+\vec{q}_1\cdot \vec{\varepsilon}(p) \vec{q}_2\cdot \vec{\varepsilon}(q_3) \times\nonumber\\
&&\Big[g_n\Big(-I_1(D^{*0},\bar{D}^{0},\bar{D}^{*0},D^{0})+I^\prime_1(D^{*0},\bar{D}^{0},D^{*0},D^{*0})\nonumber\\
&&-I_2(D^{*0},\bar{D}^{0},\bar{D}^{*0},D^{0})\Big)+g_c\Big(-I_{1}(D^{*+},D^-,D^{*-},D^{+})\nonumber\\
&&+I^\prime_1(D^{*+},D^{-},D^{*+},D^{*+})-I_2(D^{*+},D^{-},D^{*-},D^{+})\Big)\Big]\nonumber\\
&&+\vec{q}_1\cdot \vec{q}_2 \vec{\varepsilon}(p) \cdot \vec{\varepsilon}(q_3) \times\nonumber\\
&&\Big[g_n\Big(-I_1(D^{*0},\bar{D}^{0},D^{*0},D^{*0})-I_1(D^{0},\bar{D}^{*0},D^{0},D^{*0})\nonumber\\
&&-I^\prime_1(D^{0},\bar{D}^{*0},D^{0},D^{*0})-I^\prime_1(D^{*0},\bar{D}^{0},D^{*0},D^{*0})\Big)\nonumber\\
&&+g_c\Big(-I_1(D^{*+},D^{-},D^{*+},D^{*+})-I_{1}(D^{+},D^{*-},D^{-},D^{*+})\nonumber\\
&&-I^\prime_1(D^{*+},D^{-},D^{*+},D^{*+})-I^\prime_1(D^{+},D^{*-},D^{-},D^{*+})\Big)\Big]\Big\}, \label{Eq:amp2}
\end{eqnarray}
and
\begin{eqnarray}
&&\mathcal{M}_{\pi^0\pi^0\chi_{c2}}=-\frac{2\sqrt{2}}{F_\pi^2} g^2 g_1 \Big\{\epsilon_{ikl} q^i_1 q^j_2 \varepsilon^k(p)  \varepsilon^{l}_j(q_3) \nonumber\\
&&\Big[g_n I_2(D^{*0},\bar{D}^{0},\bar{D}^{*0},D^{*0})+g_c I_{2}(D^{*+},D^-,D^{*-},D^{*+})\Big]\nonumber\\
&&+\epsilon_{jkl}q^i_1 q^j_2 \varepsilon^k(p)  \varepsilon^{l}_i(q_3) \Big[g_n I_2(D^{0},\bar{D}^{*0},D^{*0},D^{*0})\nonumber\\
&&+g_c I_{2}(D^{+},D^{*-},D^{*-},D^{*+})\Big]+ \epsilon_{jkl}q^j_1 q^k_2 \varepsilon^i(p) \varepsilon^{l}_i(q_3)\nonumber\\
&&\Big[g_n \Big(-I_1(D^{0},\bar{D}^{*0},{D}^{*0},D^{*0})+I^\prime_1(D^{0},\bar{D}^{*0},D^{*0},D^{*0})\Big)\nonumber\\
&&g_c \Big(-I_1(D^{+},D^{*-},D^{*-},D^{*+})+I^\prime_1(D^{+},D^{*-},D^{*-},D^{*+})\Big]\Big\}, \nonumber\\
\label{Eq:amp3}
\end{eqnarray}
where $\varepsilon^i(p)$ and $\varepsilon^i(q_3)$ are the polarization vector of $X(3872)$ and $\chi_{c1}$, respectively. $\varepsilon^{ij}(q_3)$ is the symmetric polarization tensor for the $\chi_{c2}$. $I_1$, $I^\prime_1$, and $I_2$ are the four-point scalar integrals, where the first particle in each square bracket denotes the top left intermediate charmed meson in the corresponding diagram, and the other intermediate charmed mesons in the same diagram are listed in the square bracket in counterclockwise order along the loop. The four-point scalar integrals $I_1$, $I^\prime_1$, and $I_2$ are given in Appendix~\ref{Sec:AppendixA}, and the decay amplitudes of $X(3872)\to \pi^+ \pi^- \chi_{cJ}$ are present in Appendix~\ref{Sec:AppendixB}. Using the amplitudes in Eqs.~(\ref{Eq:amp1})-(\ref{Eq:amp3}), the decay width of $X(3872)\rightarrow \pi\pi\chi_{cJ}$ can be obtained by performing three-body phase space integral. The decay width of $X(3872)\rightarrow \pi\pi\chi_{cJ}$ is
\begin{eqnarray}
\Gamma_{X(3872)\rightarrow \pi\pi\chi_{cJ}}&=&\frac{1}{2J+1}\frac{1}{2Sm^2_X}\frac{N^2}{32\pi^3}\int^\pi_0 \int^{m_X-m_{\chi_{cJ}}}_{2m_\pi}\nonumber\\
&&\times \vec{q}^*_1 \vec{q}_3 \sin{\theta^*_1}|\mathcal{M}_{\pi\pi\chi_{cJ}}|^2 dm_{12} d\theta^*_1 \, ,
\end{eqnarray}
where $N=\sqrt{m_X m_{\chi_{cJ}}}$ accounts for the nonrelativistic normalization. Similar factors for the intermediate charmed mesons have been absorbed in the definition of the loop function. $J$ is the total spin of the initial particle. The symmetry factor $S$ is taken to be 2 for $X(3872)\to \pi^0 \pi^0 \chi_{cJ}$ decays (considering identical $\pi^0 \pi^0$ particles in the final states) and 1 for $X(3872)\to \pi^+ \pi^- \chi_{cJ}$ decays. The $|\vec{q}^*_1|$ and $|\vec{q}_3|$ are the three-momenta of the outgoing $\pi$ meson in the center of mass frame of the final $\pi\pi$ system and the outgoing $\chi_{cJ}$ meson in the $X(3872)$ rest frame, respectively. They are given by
\begin{eqnarray}
|\vec{q}_{1}^{*}|&=&\frac{\lambda^{1/2}(m_{12}^2,m_1^2,m_2^2)}{2m_{12}},\nonumber\\
|\vec{q}_{3}|&=&\frac{\lambda^{1/2}(m^2,m_{12}^2,m_3^2)}{2m},
\end{eqnarray}
where $\lambda(x,y,z)$ is the K$\ddot{\mathrm{a}}$hlen or triangle function.

\section{Numerical results and discussion}
\label{sec:results}

Before estimating the partial widths of the considered processes, we first discuss the coupling constants involved in our calculations. As a molecular state, the $X(3872)$ lies slightly below $D^* \bar{D}$ threshold. The effective coupling constant $g_{\mathrm{NR}}$, indicating the coupling between $X(3872)$ and its components, is~\cite{Weinberg:1965zz,Baru:2003qq}
\begin{eqnarray}
g^2_{\mathrm{NR}}=\lambda^2 \frac{16\pi}{\mu}\sqrt{\frac{2\epsilon}{\mu}}[1+\mathcal{O}(\sqrt{2\mu\epsilon}r)],\label{Eq:gX}
\end{eqnarray}
where $\epsilon = m_D + m_{D^*} -M_X$ denotes the binding energy of the $X(3872)$ and $\mu=m_D  m_{D^*}/(m_D + m_{D^*})$ is the reduced mass. The force range is estimated as $r^{-1}\sim\sqrt{2\mu\epsilon_{th}}$, where $\mu$ is the reduced mass and $\epsilon_{th}$ is the mass difference between the relevant threshold and the nearest one. For the $X(3872)$, this is given by $(m_{D^{*+}}+m_{D^+})-(m_{D^{*0}}+m_{D^0})$. $\lambda^2$ gives the probability of finding the molecular component in the physical state. Since the $X(3872)$ is treated as a pure molecular state, $\lambda^2=1$. Taking $m_X=3871.64$ MeV, the effective couplings are determined as $|g^n_{\mathrm{NR}}|=0.73\pm0.34\pm0.08 \ \mathrm{GeV}^{-1/2}$ and $|g^c_{\mathrm{NR}}|=2.60\pm0.03\pm1.00 \ \mathrm{GeV}^{-1/2}$, respectively, where the first uncertainty originates from the binding energies and the second stems from the approximate nature of Eq.~(\ref{Eq:gX}).

The coupling $g_\chi$ can be derived from vector meson dominance~\cite{Colangelo:2003sa,Mehen:2015efa}, given by $g_\chi=-\sqrt{{m_{\chi_{c0}}}/{6}}/{f_{\chi_{c0}}}$. With $m_{\chi_{c0}}=3417.71$ MeV and $f_{\chi_{c0}}=510$ MeV, we obtain $g_\chi=-1.48~\mathrm{GeV}^{-1/2}$. However, the vector meson dominance model may overestimate the coupling $g_\chi$ in both the charm sector~\cite{Jia:2025mtn} and the bottom sector~\cite{Guo:2016yxl}. Utilizing recent precise determinations of the pole position and the isospin-breaking properties of the $X(3872)$, the upper bound of $g^2_\chi$ is $0.28^{+1.36}_{-0.14}~\mathrm{GeV}^{-1}$~\cite{Jia:2025mtn}. Following Ref.~\cite{Jia:2025mtn}, we adopt a value of $g_\chi=0.53^{+1.82}_{-0.19}~\mathrm{GeV}^{-1/2}$. For the pionic couplings to heavy mesons, we use $g=0.79\pm0.13$ in this work, which is extracted from a tree-level calculation of the $D^{*+}$ width.

%following Ref.~\cite{Mehen:2015efa}.

\begin{table*}
\centering
\caption{The branching ratios of $X(3872)\rightarrow \pi\pi\chi_{cJ}$ ($J=0,1,2$) in the present work for different $\theta$ values  (in units of $10^{-4}$, $10^{-3}$, and $10^{-5}$ for $J=0$, $1$, and $2$, respectively). The ratio $\mathcal{R}$ is defined in Eq.~(\ref{Eq:ratioPsi}), which is in units of $10^{-3}$, $10^{-2}$, and $10^{-4}$ for $J=0$, $1$, and $2$, respectively. The branching fractions estimated by other theoretical studies~\cite{Fleming:2008yn,Fleming:2011xa,Dong:2009yp,Achasov:2024anu} and experimental measurements~\cite{BESIII:2022kow,BESIII:2023eeb,BESIII:2024ilt} are also listed for comparison.
\label{Table:br}}
\begin{ruledtabular}
\begin{tabular}{ccccccccccccc}
  $\mathrm{Channel}$ & \multicolumn{5}{c}{Present work} & Ref.~\cite{Fleming:2008yn} & Ref.~\cite{Fleming:2011xa} & Ref.~\cite{Dong:2009yp} & Ref.~\cite{Achasov:2024anu} & Exp. data~\cite{BESIII:2022kow,BESIII:2023eeb,BESIII:2024ilt} \\
  \colrule
  & 0 & $\pi/12$ & $\pi/4$ & $\pi/2$ & $\mathcal{R}$ & & & & & \\  [0.2ex]
  \cline{2-6} \\[0.2ex]
   $\pi^0\pi^0\chi_{c0}$
      & $0.17^{+0.83}_{-0.15}$  & $0.41^{+1.99}_{-0.35}$  & $0.90^{+4.39}_{-0.75}$  & $0.86^{+4.20}_{-0.69}$ & $0.40^{+1.93}_{-0.37}-2.09^{+10.2}_{-1.87}$
      & $1.4\times10^{-6}$ & ... & $0.79\times10^{-4}$ & ... & $<7.31\%$ \\[2ex]

   $\pi^+\pi^-\chi_{c0}$
      & $0.25^{+1.23}_{-0.23}$  & $0.62^{+3.03}_{-0.54}$  & $1.40^{+6.86}_{-1.17}$  & $1.38^{+6.75}_{-1.11}$ & $0.58^{+2.87}_{-0.57}-3.26^{+16.0}_{-2.92}$
      & $2.8\times10^{-6}$ & ... & ... & ... & $<2.41\%$ \\[2ex]

   $\pi^0\pi^0\chi_{c1}$
      & $0.57^{+2.79}_{-0.51}$  & $1.18^{+5.79}_{-1.03}$  & $2.30^{+11.3}_{-1.94}$  & $1.97^{+9.64}_{-1.59}$ & $1.33^{+6.50}_{-1.26}-5.35^{+26.3}_{-4.84}$
      & $2.3\%$ & $1.1\times10^{-4}$ & $0.92\times10^{-3}$ & $(0.77-1.63)\times10^{-4}$ & $<4.73\%$ \\[2ex]

   $\pi^+\pi^-\chi_{c1}$
      & $0.57^{+2.80}_{-0.51}$  & $1.37^{+6.71}_{-1.19}$  & $3.02^{+14.8}_{-2.53}$  & $2.90^{+14.2}_{-2.34}$ & $1.33^{+6.53}_{-1.26}-7.02^{+34.5}_{-6.31}$
      & $\mathcal{O}(10^{-5})$ & ... & ... & $(0.86-1.77)\times10^{-4}$ & $<0.77\%$ \\[2ex]

   $\pi^0\pi^0\chi_{c2}$
      & $0.15^{+0.76}_{-0.14}$  & $0.42^{+2.05}_{-0.36}$  & $1.01^{+4.96}_{-0.84}$  & $1.06^{+5.20}_{-0.86}$ & $0.35^{+1.77}_{-0.34}-2.47^{+12.1}_{-2.16}$
      & $3.9\times10^{-7}$ & ... & $2.55\times10^{-5}$ & ... & $<2.15\%$ \\[2ex]

   $\pi^+\pi^-\chi_{c2}$
      & $0.13^{+0.62}_{-0.11}$  & $0.35^{+1.73}_{-0.30}$  & $0.88^{+4.28}_{-0.73}$  & $0.93^{+4.58}_{-0.75}$ & $0.30^{+1.46}_{-0.27}-2.16^{+10.7}_{-1.88}$
      & $7.8\times10^{-7}$ & ... & ... & ... & -
\end{tabular}
\end{ruledtabular}
\end{table*}

In Table~\ref{Table:br}, we present the branching ratios for $X(3872)\rightarrow \pi\pi\chi_{cJ}$ as functions of the mixing angle $\theta$, considering representative values $\theta=0, \pi/2, \pi/4$, and $\pi/2$. The dominant uncertainties in our model originate from the couplings at the four vertices of the box diagrams\textemdash specifically, the errors of the coupling constants $g^{n/c}_{NR}$, $g$, and $g_{\chi}$. The angles $\theta=0$ and $\theta=\pi/2$ denote the $X(3872)$ as a pure $D^{0}\bar D^{\ast 0}$+ c.c and a pure $D^{+}D^{\ast -}$+ c.c molecular state, respectively. The angle $\theta=\pi/4$ represents an equal proportion of neutral and charged molecular components, while $\theta=\pi/12$ approximately stands for an isospin singlet state with predominantly neutral constituents. The orders of magnitude for the branching fractions of $X(3872)\rightarrow \pi\pi\chi_{cJ}$ with $J=0$, $1$, and $2$ are $10^{-4}$, $10^{-3}$, and $10^{-5}$, respectively. The channels $\pi\pi\chi_{c0/1}$ achieve the maximum at $\theta=\pi/4$, while channel $\pi\pi\chi_{c2}$ increases monotonically from $\theta=0$ to $\pi/2$. It is interesting to see that the channel $\pi^+\pi^-\chi_{c1}$ seems more sensitive to the mixing angle $\theta$ than other channels, reflecting that the decay $X(3872)\rightarrow \pi^+\pi^-\chi_{c1}$ is sensitive to the proportion of neutral and charged molecular components in the $X(3872)$.

From Eq.~(\ref{Eq:Ratio2}), one can find that the upper limits of the branching ratios of $X(3872)\rightarrow \pi^0\pi^0\chi_{c0}$ and $X(3872)\rightarrow \pi^+\pi^-\chi_{c0}$ are $7.31\%$ and $2.41\%$, respectively. Our results indicate that the corresponding branching ratios are $(0.17^{+0.83}_{-0.15}-0.90^{+4.39}_{-0.75})\times10^{-4}$ and $(0.25^{+1.23}_{-0.23}-1.40^{+6.86}_{-1.17})\times10^{-4}$, respectively, which are 2 orders of magnitude smaller than the upper limits measured by the BESIII Collaboration. In Ref.~\cite{Fleming:2008yn}, the ratio $\mathcal{B}[X(3872)\rightarrow \pi^0\pi^0\chi_{c0}]/\mathcal{B}[X(3872)\rightarrow \pi^0\chi_{c0}]$ is estimated as $9.1\times10^{-6}$ at leading order (LO) in $\chi$PT. Given the upper limit $\mathcal{B}[X(3872)\rightarrow \pi^0\chi_{c0}]=15.5\%$~\cite{BESIII:2022kow}, the upper limits are $\mathcal{B}[X(3872)\rightarrow \pi^0\pi^0\chi_{c0}]=1.4\times10^{-6}$ and $\mathcal{B}[X(3872)\rightarrow \pi^+\pi^-\chi_{c0}]\leq 2.8\times10^{-6}$ under isospin symmetry~\cite{Fleming:2008yn}, which are about 4 orders and 2 orders of magnitude smaller than the BESIII upper limits and our present estimations, respectively. In addition, in Ref.~\cite{Dong:2009yp}, the branching ratios for $X(3872)\rightarrow \pi^0\pi^0\chi_{cJ}$ are estimated to be $0.79\times10^{-4}$, $0.92\times10^{-3}$, and $2.55\times10^{-5}$ for $J=0$, $1$, and $2$, respectively. This corresponds to the case of $\theta=0$ in our work, and the values are mutually consistent.

\begin{figure*}
    \centering
    \includegraphics[width=1.0\linewidth]{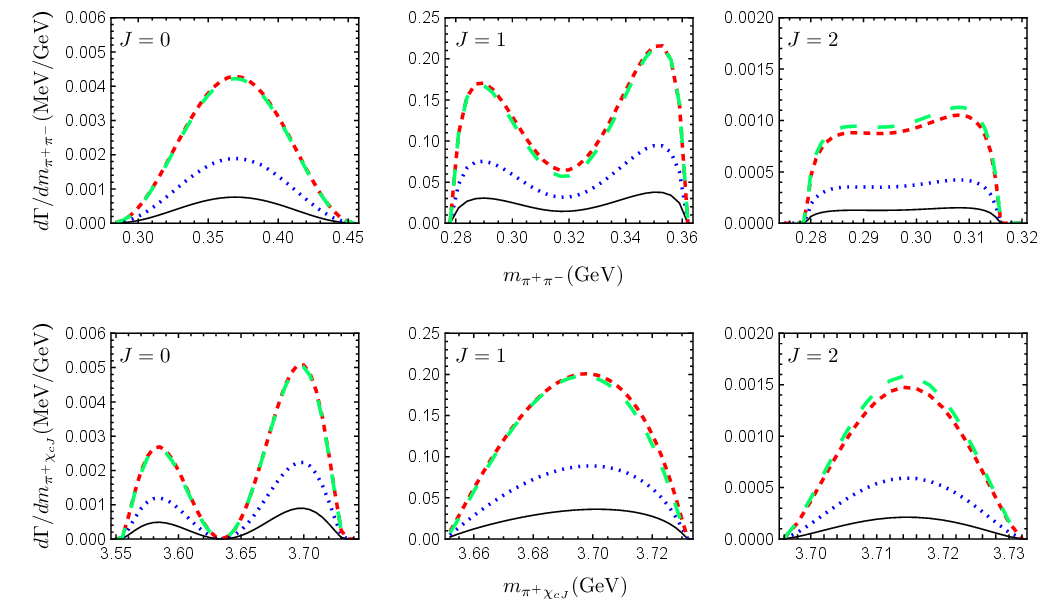}
    \caption{Invariant mass distributions of $\pi^+\pi^-$ and $\pi^+\chi_{cJ} $  for the decays of $X(3872) \to \pi^+\pi^-\chi_{cJ}\ (J=0,1,2)$. The black, blue, red, and green lines correspond to $\theta=0$, $\pi/12$, $\pi/4$, and $\pi/2$, respectively.}
    \label{fig:dis}
\end{figure*}

As for the decay processes $X(3872)\to \pi\pi\chi_{c1}$, the upper limits of the branching ratios estimated from Eq.~(\ref{Eq:Ratio3}) are $4.73\%$ and $0.77\%$ for $X(3872)\to \pi^0\pi^0\chi_{c1}$ and $X(3872)\to \pi^+\pi^-\chi_{c1}$, respectively. Our results indicate that the branching ratios of $X(3872)\to \pi^0\pi^0\chi_{c1}$ and $X(3872)\to \pi^+\pi^-\chi_{c1}$ are $(0.57^{+2.79}_{-0.51}-2.30^{+11.3}_{-1.94})\times10^{-3}$ and $(0.57^{+2.80}_{-0.51}-3.02^{+14.8}_{-2.53})\times10^{-3}$, respectively. The branching ratio for $X(3872)\to \pi^0\pi^0\chi_{c1}$  estimated in the present work is one order of magnitude smaller than the upper limits reported by the BESIII Collaboration. As for the processes $X(3872)\to \pi^+\pi^-\chi_{c1}$, our predictions for the branching ratio  is compatible with the BESIII Collaboration's upper limits. In Ref.~\cite{Fleming:2008yn}, using $\mathcal{B}[X(3872)\to \pi^0\chi_{c1}]=3.8\%$~\cite{BESIII:2019esk}, $\mathcal{B}[X(3872)\to \pi^0\pi^0\chi_{c1}]$ was estimated to be $2.32\%$, while $\mathcal{B}[X(3872)\to \pi^+\pi^-\chi_{c1}]=\mathcal{O}(10^{-5})$~\cite{Fleming:2008yn}. A later study in Ref.~\cite{Fleming:2011xa} revised $\mathcal{B}[X(3872)\to \pi^0\pi^0\chi_{c1}]$ to be $1.1\times10^{-4}$ through corrected treatment of infrared divergences. Moreover, in Ref.~\cite{Achasov:2024anu}, the branching ratios of $X(3872)\to \pi^0 \pi^0 \chi_{c1}$ and $X(3872)\to \pi^+ \pi^- \chi_{c1}$ were estimated to be $(0.8-1.7)\times10^{-4}$, which is lower than our present estimations by $1-2$ orders of magnitude and consistent with our power-counting analysis.

For the decay process $X(3872)\to \pi\pi\chi_{c2}$, the upper limit estimated from Eq.~(\ref{Eq:Ratio3}) is $2.15\%$. Our results indicate that the branching ratios of $X(3872)\rightarrow \pi^0\pi^0\chi_{c2}$ and $X(3872)\rightarrow \pi^+\pi^-\chi_{c2}$ are $(0.15^{+0.76}_{-0.14}-1.06^{+5.20}_{-0.86})\times10^{-5}$ and $(0.13^{+0.62}_{-0.11}-0.93^{+4.58}_{-0.75})\times10^{-5}$, respectively, which are 3 orders of magnitude lower than the upper limit measured from the BESIII Collaboration. Using $\mathcal{B}[X(3872)\rightarrow \pi^0\chi_{c2}]<5\%$~\cite{ParticleDataGroup:2024cfk}, isospin symmetry yields upper limits of $\mathcal{B}[X(3872)\rightarrow \pi^0\pi^0\chi_{c2}]\lesssim3.9\times10^{-7}$ and $\mathcal{B}[X(3872)\rightarrow \pi^+\pi^-\chi_{c2}]\lesssim7.8\times10^{-7}$~\cite{Fleming:2008yn}, 5 orders of magnitude below BESIII measurements.

The ratios of $\mathcal{B}[X(3872)\rightarrow \pi\pi\chi_{cJ}]$ and $\mathcal{B}[X(3872)\rightarrow \pi^+\pi^- J/\psi]$ should be more accessible experimentally. Using $\mathcal{B}[X(3872)\to \pi^+\pi^- J/\psi]=(4.3\pm1.4)\%$~\cite{ParticleDataGroup:2024cfk} and the branching ratios of $X(3872)\rightarrow \pi\pi\chi_{cJ}$ in Table~\ref{Table:br}, the ratios,
\begin{eqnarray}
\mathcal{R}=\frac{\mathcal{B}[X(3872)\rightarrow \pi\pi\chi_{cJ}]}{\mathcal{B}[X(3872)\rightarrow \pi^+\pi^- J/\psi]}, \label{Eq:ratioPsi}
\end{eqnarray}
are estimated and the values are listed in Table~\ref{Table:br}. Experimental upper limits for the ratios $\mathcal{B}[X(3872)\rightarrow \pi\pi\chi_{cJ}]/\mathcal{B}[X(3872)\rightarrow \pi^+\pi^- J/\psi]$ are given in Eqs.~(\ref{Eq:Ratio2}) and (\ref{Eq:Ratio3}), except for $\mathcal{B}[X(3872)\rightarrow \pi^+\pi^-\chi_{c2}]/\mathcal{B}[X(3872)\rightarrow \pi^+\pi^- J/\psi]$. Comparison with $\mathcal{R}$ from the present work shows that our results are lower than these experimental upper limits by $2-3$ orders of magnitude for $J=0$ and $2$. For $J=1$, the
predicted ratio $\mathcal{B}[X(3872)\rightarrow \pi^0\pi^0\chi_{c1}]/\mathcal{B}[X(3872)\rightarrow \pi^+\pi^- J/\psi]$ is an order of magnitude below the experimental upper limit, while $\mathcal{B}[X(3872)\rightarrow \pi^+\pi^-\chi_{c1}]/\mathcal{B}[X(3872)\rightarrow \pi^+\pi^- J/\psi]$ covers the experimental upper limit in the range of $\theta=0-\pi/2$. We, therefore, conclude that precision measurements of the decays $X(3872)\rightarrow \pi\pi\chi_{c1}$ are promising. Given that theoretical predictions lie $2-3$ orders of magnitude below current experimental upper limits, the decays $X(3872)\rightarrow \pi\pi\chi_{c0/2}$ require more data samples and more precise experimental studies.

Using the branching ratios in Table~\ref{Table:br}, we can also estimate the ratios for $X(3872)\rightarrow \pi\pi\chi_{cJ}$ with different $J$. For this purpose, we define
\begin{eqnarray}
R^n_{0/2}=\frac{\mathcal{B}[X(3872)\rightarrow \pi^0\pi^0\chi_{c0/2}]}{\mathcal{B}[X(3872)\rightarrow \pi^0\pi^0\chi_{c1}]},\nonumber\\
R^c_{0/2}=\frac{\mathcal{B}[X(3872)\rightarrow \pi^+\pi^-\chi_{c0/2}]}{\mathcal{B}[X(3872)\rightarrow \pi^+\pi^-\chi_{c1}]}.
\label{Eq:ratio2}
\end{eqnarray}
Our results indicate the ratios $R^n_{0}$, $R^n_{2}$, $R^c_{0}$ and $R^c_{2}$ are $(2.98^{+20.6}_{-3.75}-3.91^{+27.1}_{-4.64})\times10^{-2}$, $(2.63^{+18.5}_{-3.40}-4.61^{+32.0}_{-5.39})\times10^{-3}$, $(4.39^{+30.5}_{-5.63}-4.64^{+32.1}_{-5.49})\times10^{-2}$, and $(2.28^{+15.6}_{-2.81}-3.08^{+21.4}_{-3.58})\times10^{-3}$, respectively.
The predicted ratios can be tested by future experiments. These measurements will help us better understand the molecular nature of the $X(3872)$ and constrain the mixing angle $\theta$.

In Ref.~\cite{Achasov:2024anu}, the ratio $\mathcal{B}[X(3872)\rightarrow \pi^+\pi^-\chi_{c1}]/\mathcal{B}[X(3872)\rightarrow \pi^0\pi^0\chi_{c1}]$ was determined as $1.1$. This deviates from the expectation of 2 under exact isospin symmetry. In order to investigate the ratios of the considered decay processes with charged and neutral dipion, we define
\begin{eqnarray}
R_J=\frac{\mathcal{B}[X(3872)\rightarrow \pi^+\pi^-\chi_{cJ}]}{\mathcal{B}[X(3872)\rightarrow \pi^0\pi^0\chi_{cJ}]}.\label{Eq:ratio}
\end{eqnarray}

The ratio $R_1$ is estimated to be $1.00^{+6.93}_{-1.27}-1.31^{+9.11}_{-1.56}$, consistent with the estimations in Ref.~\cite{Achasov:2024anu}, which is $1.07-1.22$. Notably, $R_1\approx1.00$ at $\theta=0$ implies strong isospin violation in $X(3872) \to \pi\pi\chi_{c1}$ transitions for a pure neutral molecular state. As $\theta$ increases from $0$ to $\pi/2$, the ratios $R_0$ and $R_2$ are in the range of $1.47^{+10.2}_{-1.87}-1.56^{+10.8}_{-1.84}$ and $0.87^{+6.03}_{-1.09}-0.88^{+6.10}_{-1.00}$, respectively. Experimental measurement of $R_J$ could constrain $\theta$ and elucidate the $X(3872)$ molecular structure. Although the ratios $R_J$ deviate from the isospin-symmetric expectation of 2, the degree of deviation differs for $J=0,1,2$. $R_0$ and $R_1$ are closer to $2$ and $1$, respectively, and $R_2$ smaller than 1, indicating that the degree of isospin-symmetry violation follows the ordering $X(3872)\rightarrow \pi\pi\chi_{c2}>X(3872)\rightarrow \pi\pi\chi_{c1}>X(3872)\rightarrow \pi\pi\chi_{c0}$.

The ratios $R_J$ probe the isospin structure of the $X(3872)$ molecular state, which is a superposition of neutral $(|D^{*0}\bar{D}^0+\bar{D}^0D^{*0}\rangle)$ and charged $(|D^{*+}D^{-}+D^{+}D^{*-}\rangle)$ components parametrized by a mixing angle $\theta$. These ratios provide crucial insight into the isospin configuration of $X(3872)$. The ratios $R_0,R_1$ and $R_2$ vary with $\theta$, demonstrating how isospin mixing influences decay patterns. The ratios $R_J$ exhibit minimal $\theta$ dependence, increasing slightly with $\theta$. This indicates that the charged components enhance the decays $X(3872)\rightarrow \pi^+\pi^-\chi_{cJ}$. Deviations of $R_J$ from $2$ reveal significant isospin-symmetry violation in $X(3872)\rightarrow \pi\pi\chi_{cJ}$ decays. When masses of the isospin-triplet $D^{(*)}$ and $\pi$ states are set equal, i.e., $m_{D^{(*)+}}=m_{D^{(*)0}}$ and $m_{\pi^\pm}=m_{\pi^0}$, the ratios $R_J$ are equal to $2$ at $\theta=\pi/12$. Thus, isospin breaking in $X(3872) \to \pi\pi\chi_{cJ}$ originates from the mass difference between the $u$ and $d$ quarks.

In addition to calculating the branching ratios for the decays $X(3872) \to \pi^+ \pi^- \chi_{cJ}\ (J=0,1,2)$, we have also analyzed the invariant mass distributions of both the $ \pi^+\pi^- $ system and the $ \pi^+\chi_{cJ} $ combination, as shown in Fig.~\ref{fig:dis}. For $J=0$, the distribution of $d\Gamma_{X\to \pi^+\pi^-\chi_{c0}}/dm_{\pi^+\pi^-}$ exhibits a broad single bump, peaking at approximately $0.369$ GeV. Interestingly, we find a double-bump structure in the $d\Gamma_{X\to \pi^+\pi^-\chi_{c0}}/dm_{\pi^+\chi_{c0}}$ distribution. We suggest experimental studies of the $\pi^+\chi_{c0}$ invariant mass spectrum in the decay $X(3872)\to \pi^+\pi^-\chi_{c0}$ around $3.585$ and $3.7$ GeV. For $J=1$, a double-bump structure, which peaks at $0.29$ and $0.35$ GeV, emerges in the $d\Gamma_{X\to \pi^+\pi^-\chi_{c1}}/dm_{\pi^+\pi^-}$ distribution. The double-bump structure, resulting from loop diagrams, can be recognized as evidence for the hadronic molecular nature of the $X(3872)$ and has been proposed as an observable to distinguish the hadronic molecular and the compact state pictures of the $D_{s1}(2460)$~\cite{Tang:2023yls}. The distributions of $d\Gamma_{X\to \pi^+\pi^-\chi_{c1}}/dm_{\pi^+\chi_{c1}}$ exhibits a broad single bump and peaking at $3.702$ GeV.~\footnote{The peak of the $d\Gamma_{X\to \pi^+\pi^-\chi_{c1}}/dm_{\pi^+\chi_{c1}}$ distribution has a shift as $\theta$ increases from $0$ to $\pi/2$. The peak positions are located at $m_{\pi^+\chi_{c1}}=3.702$, $3.7$, $3.698$, and $3.696$ GeV for $\theta=0$, $\pi/12$, $\pi/4$, and $\pi/2$, respectively.} For $J=2$, the distribution of $d\Gamma_{X\to \pi^+\pi^-\chi_{c2}}/dm_{\pi^+\pi^-}$ shows a much broader bump. The distribution of $d\Gamma_{X\to \pi^+\pi^-\chi_{c2}}/dm_{\pi^+\chi_{c2}}$ shows a broad single bump and peaking at $3.714$ GeV.

\section{SUMMARY}
\label{sec:summary}

The $X(3872)$, discovered in 2003, challenges conventional quark models due to its proximity to the $D^0\bar{D}^{*0}$ threshold, suggesting a dominant molecular structure. Its decays exhibit significant isospin violation, evidenced by comparable branching ratios $\mathcal{B}(X \to \rho^0 J/\psi) - \mathcal{B}(X \to \omega J/\psi)$. Recent BESIII measurements of $\mathcal{B}(X \to \pi^0\chi_{c1}) \sim 3.8\%$ and upper limits for dipionic transitions [e.g., $\mathcal{B}(X \to \pi^+\pi^-\chi_{c1}) < 0.77\%$] motivate theoretical studies of dipionic transitions $X(3872) \to \pi\pi\chi_{cJ}$ ($J=0,1,2$) as probes of its internal structure.
	
Using HH$\chi$PT, we investigate $X(3872) \to \pi\pi\chi_{cJ}$ by treating the $X(3872)$ as a superposition of neutral and charged molecular components with a mixing angle. We employ NREFT power-counting rules to analyze box and triangle diagrams and find that box diagrams dominate the dipionic transitions $X(3872) \to \pi\pi\chi_{cJ}$. The branching ratios of $X(3872)\rightarrow \pi\pi\chi_{cJ}$ with $J=0$, $1$, and $2$ are estimated to be of the order of $10^{-4}$, $10^{-3}$ and $10^{-5}$, respectively. The ratios ${\mathcal{B}[X\to  \pi\pi\chi_{cJ}]}/{\mathcal{B}[X\to \pi^+ \pi^- J/\psi]}$ are predicted, and the results reveal that $X(3872) \to \pi\pi\chi_{c1}$ decays are prime targets for the BESIII and Belle II Collaborations, while the $\chi_{c0/2}$ channels require higher precision.

The ratios ${\mathcal{B}[X(3872)\rightarrow \pi\pi\chi_{c0/2}]}/{\mathcal{B}[X(3872)\rightarrow \pi\pi\chi_{c1}]}$ and  ${\mathcal{B}[X(3872)\rightarrow \pi^+\pi^-\chi_{cJ}]}/{\mathcal{B}[X(3872)\rightarrow \pi^0\pi^0\chi_{cJ}]}$ are also predicted. The latter ratios deviate from isospin-symmetric expectations. Specifically, the ratios are $1.47^{+10.2}_{-1.87}-1.56^{+10.8}_{-1.84}$, $1.00^{+6.93}_{-1.27}-1.31^{+9.11}_{-1.56}$, and $0.87^{+6.03}_{-1.09}-0.88^{+6.10}_{-1.00}$ for $J=0$, $1$, and $2$, respectively, showing different degrees of isospin violation and the degree of isospin-symmetry violation follows the order $X(3872)\rightarrow \pi\pi\chi_{c2}>X(3872)\rightarrow \pi\pi\chi_{c1}>X(3872)\rightarrow \pi\pi\chi_{c0}$. Future measurements of $R_J$ will constrain the mixing angle $\theta$, resolving the neutral-charged mixing fraction and refining molecular models. The invariant mass distributions of $\pi^+\pi^-$ and $\pi^+\chi_{cJ}$ for the decay of $X(3872) \to \pi^+\pi^-\chi_{cJ}$ are studied. We find a double-bump structure in the $d\Gamma_{X\to \pi^+\pi^-\chi_{c1}}/dm_{\pi^+\pi^-}$ distribution and the $d\Gamma_{X\to \pi^+\pi^-\chi_{c0}}/dm_{\pi^+\chi_{c0}}$ distribution, which could be tested by future experiments.

\section*{ACKNOWLEDGMENTS}
 Q.W. thanks Zhao-Sai Jia for useful discussions. This work is supported by the National Natural Science Foundation of China under Grants No. 12175037, No.12335001, No.12475081, No.12105153, and No.12405093, as well as supported, in part, by National Key Research and Development Program under Grants No. 2024YFA1610503 and No.2024YFA1610504. It is also supported by Taishan Scholar Project of Shandong Province under Grant No. tsqn202103062 and by the Natural Science Foundation of Shandong Province under Grants No. ZR2025MS04 and No.ZR2022ZD26.

\onecolumngrid
\section*{DATA AVAILABILITY}

The data that support the findings of this article are not publicly available. The data are available from the authors upon reasonable request.
\appendix
\section{DECAY AMPLITUDES}
\label{Sec:AppendixA}
%\begin{appendix}
%\section{Decay amplitudes}\label{appendix-A}

The decay amplitudes for the $X(3872)\to \pi^+ \pi^- \chi_{cJ}$ read
\begin{align}
%% A1
\mathcal{M}_{\pi^+\pi^-\chi_{c0}}&=\sqrt{\frac23}\frac1{F_\pi^2} g^2 g_1 q_1^i q_2^j \varepsilon^k(p) \epsilon_{i j k}
\times\Big[g_n\Big(I_{1}(\bar{D}^0,{D}^{*0},{D}^{*0},D^{*-})+3I_{1}(\bar{D}^{*0},{D}^{0},{D}^{0},D^{*-}) \nonumber\\
&-3I_{1}^{\prime}(D^{*0},\bar{D}^{0},\bar{D}^{0},D^{*+})-I_1^\prime(D^0,\bar{D}^{*0},\bar{D}^{*0},D^{*+})
+I_2(\bar{D}^{*0},{D}^0,{D}^{*+},D^{*-})-I_{2}(\bar{D}^0,{D}^{*0},{D}^{*+},D^{*-})\Big)\nonumber\\
&+g_c\Big(I_{1}(D^{+},D^{*-},D^{*-},D^{*0})+3I_{1}(D^{*+},D^-,D^-,D^{*0})
-3I_{1}^{\prime}(D^{*-},D^+,D^+,\bar{D}^{*0})-I_1^\prime(D^{-},D^{*+},D^{*+},\bar{D}^{*0})\nonumber\\
&+I_2(D^{*+},D^{-},\bar{D}^{*0},D^{*0})-I_{2}(D^{+},D^{*-},\bar{D}^{*0},D^{*0})\Big)\Big],
\end{align}
%% A2
\begin{align}
\mathcal{M}_{\pi^+\pi^-\chi_{c1}}&=\frac{2}{F_\pi^2} g^2 g_1 \Big\{\vec{q}_2\cdot \vec{\varepsilon}(p) \vec{q}_1\cdot \vec{\varepsilon}(q_3)\times
\Big[g_n\Big(I_1(\bar{D}^{*0},{D}^{0},{D}^{*0},D^{*-})-I^\prime_1(D^{*0},\bar{D}^{0},\bar{D}^{*0},D^{+})
-I_2(\bar{D}^{0},{D}^{*0},D^{+},D^{*-})\Big)\nonumber\\
&+g_c\Big(I_{1}(D^{*+},D^-,D^{*-},D^{*0})
-I^\prime_1(D^{*-},D^+,D^{*+},\bar{D}^{0})-I_2(D^{+},D^{*-},\bar{D}^0,D^{*0})\Big)\Big]\nonumber\\
&+\vec{q}_1\cdot \vec{\varepsilon}(p) \vec{q}_2\cdot \vec{\varepsilon}(q_3) \times
\Big[g_n\Big(-I_1(\bar{D}^{*0},{D}^{0},{D}^{*0},D^{-})+I^\prime_1(D^{*0},\bar{D}^{0},\bar{D}^{*0},D^{*+})
-I_2(\bar{D}^{*0},{D}^{0},{D}^{*+},D^{-})\Big)\nonumber\\
&+g_c\Big(-I_{1}(D^{*+},D^-,D^{*-},D^{0})
+I^\prime_1(D^{*-},D^{+},D^{*+},\bar{D}^{*0})-I_2(D^{*+},D^{-},\bar{D}^{*0},D^{0})\Big)\Big]\nonumber\\
&+\vec{q}_1\cdot \vec{q}_2 \vec{\varepsilon}(p) \cdot \vec{\varepsilon}(q_3) \times\Big[g_n\Big(-I_1(\bar{D}^{0},{D}^{*0},D^{0},D^{*-})-I_1(\bar{D}^{*0},{D}^{0},D^{*0},D^{*-})
-I^\prime_1(D^{0},\bar{D}^{*0},\bar{D}^{0},D^{*+})-I^\prime_1(D^{*0},\bar{D}^{0},\bar{D}^{*0},D^{*+})\Big)\nonumber\\
&+g_c\Big(-I_1(D^{*+},D^{-},D^{*-},D^{*0})-I_{1}(D^{+},D^{*-},D^{-},D^{*0})-I^\prime_1(D^{*-},D^{+},D^{*+},\bar{D}^{*0})-I^\prime_1(D^{-},D^{*+},D^{+},\bar{D}^{*0})\Big)\Big]\Big\},
\end{align}
%% A3
\begin{align}
\mathcal{M}_{\pi^+\pi^-\chi_{c2}}&=-\frac{2\sqrt{2}}{F_\pi^2} g^2 g_1 \Big\{\epsilon_{ikl}q^i_1 q^j_2 \varepsilon^k(p)  \varepsilon^{l}_j(q_3)
\Big[g_n I_2(\bar{D}^{*0},{D}^{0},{D}^{*+},D^{*-})+g_c I_{2}(D^{*+},D^-,\bar{D}^{*0},D^{*0})\Big]\nonumber\\
&+\epsilon_{jkl} q^i_1 q^j_2 \varepsilon^k(p)  \varepsilon^{l}_i(q_3) \Big[g_n I_2(\bar{D}^{0},{D}^{*0},D^{*+},D^{*-})
+g_c I_{2}(D^{+},D^{*-},\bar{D}^{*0},D^{*0})\Big]+\epsilon_{jkl}q^j_1 q^k_2 \varepsilon^i(p)  \varepsilon^{l}_i(q_3)
\Big[g_n \Big(-I_1(\bar{D}^{0},D^{*0},D^{*0},D^{*-})\nonumber\\
&+I^\prime_1(D^{0},\bar{D}^{*0},\bar{D}^{*0},D^{*+})\Big)+
g_c \Big(-I_1({D}^{+},D^{*-},D^{*-},D^{*0})+I^\prime_1(D^{-},D^{*+},D^{*+},\bar{D}^{*0})\Big)\Big]\Big\}.
\end{align}

\section{FOUR-POINT LOOP INTEGRALS}
\label{Sec:AppendixB}

\begin{figure}
    \centering    \includegraphics[width=0.95\linewidth]{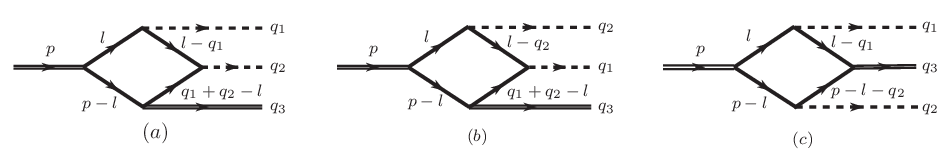}
    \caption{Kinematics used for calculating four-point integrals.}
    \label{fig:feyndiagsx}
\end{figure}

Following Refs.~\cite{Chen:2019gfp,Jia:2023pud}, we present the analytical expression of scalar four-point loop integrals. In the rest frame of the decay particle $p=(M,0)$, they are written as
\begin{align}
I_1[m_1,m_2,m_3,m_4]&=\frac{-\mu_{12}\mu_{23}\mu_{24}}{2}\int\frac{d^{3}l}{(2\pi)^{3}}\frac{1}{\left[\vec{l}^{2}+c_{12}-i\epsilon\right]
\left[\vec{l}^{2}+2\frac{\mu_{23}}{m_{3}}\vec{l}\cdot\vec{q}_{3}+c_{23}-i\epsilon\right]\left[\vec{l}^{2}-2\frac{\mu_{24}}{m_{4}}\vec{l}\cdot\vec{q}_{1}
+c_{24}-i\epsilon\right]},\label{eq:I1}\\
I_1^{\prime}[m_1,m_2,m_3,m_4]&=\frac{-\mu_{12}\mu_{23}\mu_{24}}{2}\int\frac{d^3l}{(2\pi)^3}\frac{1}{[\vec{l}^2+c_{12}-i\epsilon]
\left[\vec{l}^2+2\frac{\mu_{23}}{m_3}\vec{l}\cdot\vec{q}_3+c_{23}-i\epsilon\right]\left[\vec{l}^2-2\frac{\mu_{24}}{m_4}\vec{l}\cdot\vec{q}_2+
c_{24}^{\prime}-i\epsilon\right]},\label{eq:I1p}\\
I_{2}[m_{1},m_{2},m_{3},m_{4}]&=\frac{-\mu_{12}\mu_{34}}{2}\int\frac{d^3l}{(2\pi)^3}\frac{1}{[\vec{l}^2+c_{12}-i\epsilon]
[\vec{l}^2-\frac{2\mu_{34}}{m_4}\vec{l}\cdot\vec{q}_1+\frac{2\mu_{34}}{m_3}\vec{l}\cdot\vec{q}_2+c_{34}-i\epsilon]}\nonumber\\
& \times\left[\frac{\mu_{24}}{[\vec{l}^2-
\frac{2\mu_{24}}{m_4}\vec{l}\cdot\vec{q}_1+c_{24}-i\epsilon]}+\frac{\mu_{13}}{[\vec{l}^2+\frac{2\mu_{13}}{m_3}\vec{l}\cdot\vec{q}_2+c_{13}-i\epsilon]}\right].
\label{eq:I2}
\end{align}
The parameters $c^{(\prime)}_{ij}$ in Eqs. \eqref{eq:I1}-\eqref{eq:I2} are defined as
\begin{align}
c_{12}&\equiv2\mu_{12}(m_1+m_2-M),\\
c_{23}&\equiv2\mu_{23}\left(m_2+m_3-M+q_1^0+q_2^0+\frac{\vec{q}_3^2}{2m_3}\right),\\
c_{24}&\equiv2\mu_{24}\left(m_2+m_4-M+q_1^0
+\frac{\vec{q}_1^2}{2m_4}\right),\\
c_{24}^{\prime}&\equiv2\mu_{24}\left(m_2+m_4-M+q_2^0+\frac{\vec{q}_2^2}{2m_4}\right)\\
c_{34}&\equiv2\mu_{34}\left(m_3+m_4-q_3^0+\frac{\vec{q}_1^2}{2m_4}+\frac{\vec{q}_2^2}{2m_3}\right),\\ c_{13}&\equiv2\mu_{13}\left(m_1+m_3-M+q_2^0+\frac{\vec{q}_2^2}{2m_3}\right).
\end{align}
Note that the factor $m_{1}m_{2}m_{3}m_{4}$ has canceled with the nonrelativistic normalization factors from the intermediate charmed mesons, which is different from Refs.~\cite{Chen:2019gfp,Jia:2023pud}.

\section{CUTOFFIN SCALLAR FOUR-POINT LOOP INTEGRALS}
\label{Sec:AppendixC}

In this appendix, we discuss the parametrization and simplification of the scalar four-point integrals in the box diagrams. Take the integral in Eq.~(\ref{eq:I1}) as an example, which is
\begin{align}
&\int\frac{d^{3}l}{(2\pi)^{3}}\frac{1}{\left[\vec{l}^{2}+c_{12}-i\epsilon\right]
\left[\vec{l}^{2}+2\frac{\mu_{23}}{m_{3}}\vec{l}\cdot\vec{q}_{3}+c_{23}-i\epsilon\right]\left[\vec{l}^{2}-2\frac{\mu_{24}}{m_{4}}\vec{l}\cdot\vec{q}_{1}
+c_{24}-i\epsilon\right]}\nonumber\\
&=\frac{1}{(2\pi)^3}\int_0^{2\pi}\mathrm{d}\phi\int_{-1}^1\mathrm{d}x\int_0^{\Lambda}\mathrm{d}l\frac{l^2}{[l^2+c_{12}-i\epsilon]
[{l}^2+2\frac{\mu_{23}}{m3}\vec{l}\cdot\vec{q}_3+c_{23}-i\epsilon][-2\frac{\mu_{24}}{m_4}lq_1][x-x_0+i\epsilon]}
\end{align}
where $x_0=({l}^2+c_{24})/(2\frac{\mu_{24}}{m_4}lq_1)$ and $\Lambda$ is a cutoff.

\begin{figure*}[!htbp]
    \centering    \includegraphics[width=0.4\linewidth]{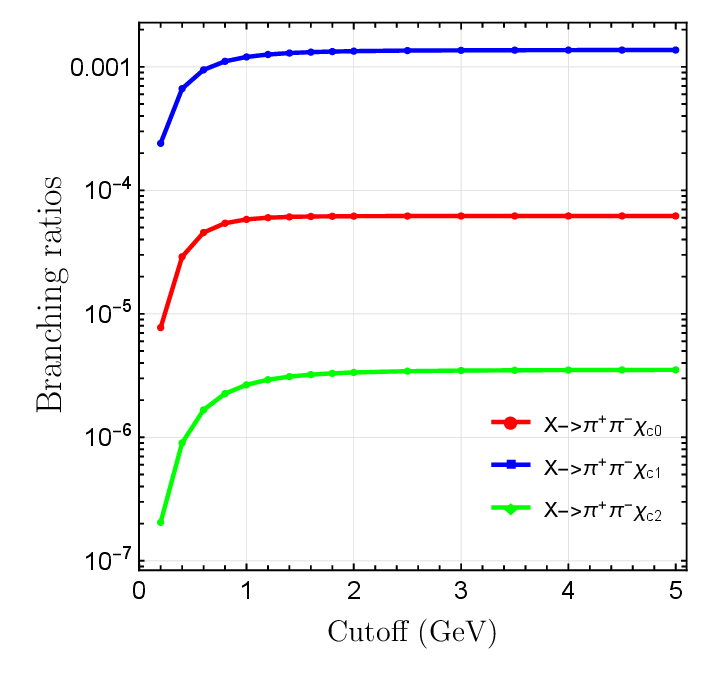}
    \caption{The cutoff dependence of the branching ratios of $X(3872) \to \pi^+ \pi^- \chi_{cJ} \ (J=0,1,2)$ when we take $\theta=\pi/12$.}
    \label{fig:cutoff}
\end{figure*}

From Fig.~\ref{fig:cutoff}, the branching ratios of $X(3872) \to \pi^+ \pi^- \chi_{cJ}$ ($J=0,1,2$) increase rapidly with the cutoff $\Lambda$ below 1.5~GeV, reflecting significant low-momentum contributions. This scale coincides with the chiral symmetry-breaking scale $\Lambda_\chi \approx 4\pi f_\pi = 1.66$~GeV in HH$\chi$PT. For $\Lambda > 1.5$~GeV, the growth slows and stabilizes, supporting the robustness of our approach. In Refs.~\cite{Fleming:2008yn,Fleming:2011xa}, the decays $X(3872) \to \chi_{cJ}(\pi^0,\pi\pi)$ were studied using HH$\chi$PT and XEFT with the power divergence subtraction (PDS) scheme, reporting only the ratios $\frac{\mathrm{Br}[X(3872) \to \chi_{cJ}\pi\pi]}{\mathrm{Br}[X(3872) \to \chi_{cJ}\pi]}$ without specifying the cutoff $\Lambda_{\mathrm{PDS}}$. Our work employs HH$\chi$PT with a hard momentum cutoff $\Lambda$, which generally differs from the PDS scheme by a factor.
Using the upper limits of $\mathrm{Br}[X(3872) \to \chi_{c1/2}\pi^0]$ estimated from Ref.~\cite{Jia:2025mtn}, we estimate the ratios $\frac{\rm{Br}[X(3872)\to \chi_{cJ}\pi^+\pi^-]}{\rm{Br}[X(3872)\to \chi_{cJ}\pi^0]}$ to be $2.28^{+11.2}_{-1.98}\times10^{-2}$ and $7.00^{+34.6}_{-6.00}\times10^{-5}$ for $J=1$ and $2$, respectively. Our results are approximately a factor of 4 larger than those in Refs.~\cite{Fleming:2008yn,Fleming:2011xa}. This corresponds to $\Lambda \approx 300$~MeV in our calculation.  The factorization approach is limited by the charged channel binding momentum $\gamma_c \approx 126$~MeV. For $\Lambda \gg \gamma_c$, the wave function becomes unreliable in high-momentum regions, preventing recovery of factorization results. The estimated value $\Lambda \approx 300$~MeV is of the same order as $\gamma_c$ and, thus, remains reasonable.

\twocolumngrid

\bibliographystyle{unsrt}
\bibliography{references}

@article{Belle:2003nnu,
    author = "Choi, S. K. and others",
    collaboration = "Belle Collaboration",
    title = "{Observation of a narrow charmonium-like state in exclusive $B^\pm \to K^\pm \pi^+ \pi^- J/\psi$ decays}",
    archivePrefix = "arXiv",
    doi = "10.1103/PhysRevLett.91.262001",
    journal = "Phys. Rev. Lett.",
    volume = "91",
    pages = "262001",
    year = "2003"
}

@article{Chen:2016qju,
  author =        {Chen, Hua-Xing and Chen, Wei and Liu, Xiang and
                   Zhu, Shi-Lin},
  journal =       {Phys. Rep.},
  pages =         {1--121},
  title =         {{The hidden-charm pentaquark and tetraquark states}},
  volume =        {639},
  year =          {2016},
  doi =           {10.1016/j.physrep.2016.05.004},
}

@article{Hosaka:2016pey,
  author =        {Hosaka, Atsushi and Iijima, Toru and
                   Miyabayashi, Kenkichi and Sakai, Yoshihide and
                   Yasui, Shigehiro},
  journal =       {Prog.Theor.Exp.Phys.},
  number =        {6},
  pages =         {062C01},
  title =         {{Exotic hadrons with heavy flavors: X, Y, Z, and
                   related states}},
  volume =        {2016},
  year =          {2016},
  doi =           {10.1093/ptep/ptw045},
}

@article{Lebed:2016hpi,
  author =        {Lebed, Richard F. and Mitchell, Ryan E. and
                   Swanson, Eric S.},
  journal =       {Prog. Part. Nucl. Phys.},
  pages =         {143--194},
  title =         {{Heavy-Quark QCD Exotica}},
  volume =        {93},
  year =          {2017},
  doi =           {10.1016/j.ppnp.2016.11.003},
}

@article{Esposito:2016noz,
  author =        {Esposito, A. and Pilloni, A. and Polosa, A. D.},
  journal =       {Phys. Rep.},
  pages =         {1--97},
  title =         {{Multiquark Resonances}},
  volume =        {668},
  year =          {2017},
  doi =           {10.1016/j.physrep.2016.11.002},
}

@article{Guo:2017jvc,
  author =        {Guo, Feng-Kun and Hanhart, Christoph and
                   Mei\ss{}ner, Ulf-G. and Wang, Qian and Zhao, Qiang and
                   Zou, Bing-Song},
  journal =       {Rev. Mod. Phys.},
  note =          {\textbf{94}, 029901(E)(2022)},
  number =        {1},
  pages =         {015004},
  title =         {{Hadronic molecules}},
  volume =        {90},
  year =          {2018},
  doi =           {10.1103/RevModPhys.90.015004},
}

@article{Ali:2017jda,
  author =        {Ali, Ahmed and Lange, Jens S\"oren and
                   Stone, Sheldon},
  journal =       {Prog. Part. Nucl. Phys.},
  pages =         {123--198},
  title =         {{Exotics: Heavy Pentaquarks and Tetraquarks}},
  volume =        {97},
  year =          {2017},
  doi =           {10.1016/j.ppnp.2017.08.003},
}

@article{Olsen:2017bmm,
  author =        {Olsen, Stephen Lars and Skwarnicki, Tomasz and
                   Zieminska, Daria},
  journal =       {Rev. Mod. Phys.},
  number =        {1},
  pages =         {015003},
  title =         {{Nonstandard heavy mesons and baryons: Experimental
                   evidence}},
  volume =        {90},
  year =          {2018},
  doi =           {10.1103/RevModPhys.90.015003},
}

@article{Karliner:2017qhf,
  author =        {Karliner, Marek and Rosner, Jonathan L. and
                   Skwarnicki, Tomasz},
  journal =       {Annu. Rev. Nucl. Part. Sci.},
  pages =         {17--44},
  title =         {{Multiquark States}},
  volume =        {68},
  year =          {2018},
  doi =           {10.1146/annurev-nucl-101917-020902},
}

@article{Yuan:2018inv,
  author =        {Yuan, Chang-Zheng},
  journal =       {Int. J. Mod. Phys. A},
  number =        {21},
  pages =         {1830018},
  title =         {{The XYZ states revisited}},
  volume =        {33},
  year =          {2018},
  doi =           {10.1142/S0217751X18300181},
}

@article{Dong:2017gaw,
  author =        {Dong, Yubing and Faessler, Amand and
                   Lyubovitskij, Valery E.},
  journal =       {Prog. Part. Nucl. Phys.},
  pages =         {282--310},
  title =         {{Description of heavy exotic resonances as molecular
                   states using phenomenological Lagrangians}},
  volume =        {94},
  year =          {2017},
  doi =           {10.1016/j.ppnp.2017.01.002},
}

@article{Liu:2019zoy,
  author =        {Liu, Yan-Rui and Chen, Hua-Xing and Chen, Wei and
                   Liu, Xiang and Zhu, Shi-Lin},
  journal =       {Prog. Part. Nucl. Phys.},
  pages =         {237--320},
  title =         {{Pentaquark and Tetraquark states}},
  volume =        {107},
  year =          {2019},
  doi =           {10.1016/j.ppnp.2019.04.003},
}

@article{Liu:2024uxn,
    author = "Liu, Ming-Zhu and Pan, Ya-Wen and Liu, Zhi-Wei and Wu, Tian-Wei and Lu, Jun-Xu and Geng, Li-Sheng",
    journal = {Phys.Rep.\textbf{1108},1},
    title = "{Three ways to decipher the nature of exotic hadrons: multiplets, three-body hadronic molecules, and correlation functions}",
    archivePrefix = "arXiv",
    primaryClass = "hep-ph",
    month = "4",
    year = "2025"
}

@article{LHCb:2015jfc,
    author = "Aaij, Roel and others",
    collaboration = "LHCb Collaboration",
    title = "{Quantum numbers of the $X(3872)$ state and orbital angular momentum in its $\rho^0 J\psi$ decay}",
    archivePrefix = "arXiv",
    primaryClass = "hep-ex",
    reportNumber = "LHCB-PAPER-2015-015, CERN-PH-EP-2015-098",
    doi = "10.1103/PhysRevD.92.011102",
    journal = "Phys. Rev. D",
    volume = "92",
    number = "1",
    pages = "011102",
    year = "2015",
}

@article{Godfrey:1985xj,
    author = "Godfrey, S. and Isgur, Nathan",
    title = "{Mesons in a Relativized Quark Model with Chromodynamics}",
    doi = "10.1103/PhysRevD.32.189",
    journal = "Phys. Rev. D",
    volume = "32",
    pages = "189--231",
    year = "1985"
}

@article{Tornqvist:1993ng,
    author = "Tornqvist, Nils A.",
    title = "{From the deuteron to deusons, an analysis of deuteron - like meson meson bound states}",
    archivePrefix = "arXiv",
    reportNumber = "HU-SEFT-R-1993-12",
    doi = "10.1007/BF01413192",
    journal = "Z. Phys. C",
    volume = "61",
    pages = "525--537",
    year = "1994"
}

@article{Voloshin:2003nt,
    author = "Voloshin, M. B.",
    title = "{Interference and binding effects in decays of possible molecular component of X(3872)}",
    archivePrefix = "arXiv",
    reportNumber = "TPI-MINN-03-27-T, UMN-TH-2216-03",
    doi = "10.1016/j.physletb.2003.11.014",
    journal = "Phys. Lett. B",
    volume = "579",
    pages = "316--320",
    year = "2004"
}

@article{Swanson:2003tb,
    author = "Swanson, Eric S.",
    title = "{Short range structure in the X(3872)}",
    archivePrefix = "arXiv",
    reportNumber = "JLAB-THY-03-227",
    doi = "10.1016/j.physletb.2004.03.033",
    journal = "Phys. Lett. B",
    volume = "588",
    pages = "189--195",
    year = "2004"
}

@article{Tornqvist:2004qy,
    author = "Tornqvist, Nils A.",
    title = "{Isospin breaking of the narrow charmonium state of Belle at 3872-MeV as a deuson}",
    archivePrefix = "arXiv",
    doi = "10.1016/j.physletb.2004.03.077",
    journal = "Phys. Lett. B",
    volume = "590",
    pages = "209--215",
    year = "2004"
}

@article{Fleming:2007rp,
    author = "Fleming, S. and Kusunoki, M. and Mehen, T. and van Kolck, U.",
    title = "{Pion interactions in the $X(3872)$}",
    archivePrefix = "arXiv",
    reportNumber = "JLAB-THY-07-615",
    doi = "10.1103/PhysRevD.76.034006",
    journal = "Phys. Rev. D",
    volume = "76",
    pages = "034006",
    year = "2007"
}

@article{Liu:2008fh,
    author = "Liu, Yan-Rui and Liu, Xiang and Deng, Wei-Zhen and Zhu, Shi-Lin",
    title = "{Is $X(3872) $ Really a Molecular State?}",
    archivePrefix = "arXiv",
    primaryClass = "hep-ph",
    doi = "10.1140/epjc/s10052-008-0640-4",
    journal = "Eur. Phys. J. C",
    volume = "56",
    pages = "63--73",
    year = "2008"
}

@article{Swanson:2004pp,
    author = "Swanson, Eric S.",
    title = "{Diagnostic decays of the X(3872)}",
    archivePrefix = "arXiv",
    doi = "10.1016/j.physletb.2004.07.059",
    journal = "Phys. Lett. B",
    volume = "598",
    pages = "197--202",
    year = "2004"
}

@article{Braaten:2003he,
    author = "Braaten, Eric and Kusunoki, Masaoki",
    title = "{Low-energy universality and the new charmonium resonance at 3870-MeV}",
    archivePrefix = "arXiv",
    doi = "10.1103/PhysRevD.69.074005",
    journal = "Phys. Rev. D",
    volume = "69",
    pages = "074005",
    year = "2004"
}

@article{Braaten:2005ai,
    author = "Braaten, Eric and Kusunoki, Masaoki",
    title = "{Decays of the X(3872) into J/psi and light hadrons}",
    archivePrefix = "arXiv",
    doi = "10.1103/PhysRevD.72.054022",
    journal = "Phys. Rev. D",
    volume = "72",
    pages = "054022",
    year = "2005"
}

@article{Braaten:2004ai,
    author = "Braaten, Eric and Kusunoki, Masaoki",
    title = "{Exclusive production of the X(3872) in B meson decay}",
    archivePrefix = "arXiv",
    doi = "10.1103/PhysRevD.71.074005",
    journal = "Phys. Rev. D",
    volume = "71",
    pages = "074005",
    year = "2005"
}

@article{Liu:2006df,
    author = "Liu, Xiang and Zhang, Bo and Zhu, Shi-Lin",
    title = "{The Hidden Charm Decay of X(3872), Y(3940) and Final State Interaction Effects}",
    archivePrefix = "arXiv",
    doi = "10.1016/j.physletb.2006.12.031",
    journal = "Phys. Lett. B",
    volume = "645",
    pages = "185--188",
    year = "2007"
}

@article{Dubynskiy:2007tj,
    author = "Dubynskiy, S. and Voloshin, Mikhail B.",
    title = "{Pionic transitions from $X(3872) $ to chi(cJ)}",
    archivePrefix = "arXiv",
    primaryClass = "hep-ph",
    reportNumber = "FTPI-MINN-07-29, UMN-TH-2619-07",
    doi = "10.1103/PhysRevD.77.014013",
    journal = "Phys. Rev. D",
    volume = "77",
    pages = "014013",
    year = "2008"
}

@article{Fleming:2008yn,
    author = "Fleming, Sean and Mehen, Thomas",
    title = "{Hadronic Decays of the X(3872) to chi(cJ) in Effective Field Theory}",
    archivePrefix = "arXiv",
    primaryClass = "hep-ph",
    doi = "10.1103/PhysRevD.78.094019",
    journal = "Phys. Rev. D",
    volume = "78",
    pages = "094019",
    year = "2008"
}

@article{Dong:2008gb,
    author = "Dong, Yu-bing and Faessler, Amand and Gutsche, Thomas and Lyubovitskij, Valery E.",
    title = "{Estimate for the X(3872) ---{\ensuremath{>}} gamma J/psi decay width}",
    archivePrefix = "arXiv",
    primaryClass = "hep-ph",
    doi = "10.1103/PhysRevD.77.094013",
    journal = "Phys. Rev. D",
    volume = "77",
    pages = "094013",
    year = "2008"
}

@article{Bignamini:2009sk,
    author = "Bignamini, C. and Grinstein, B. and Piccinini, F. and Polosa, A. D. and Sabelli, C.",
    title = "{Is the X(3872) Production Cross Section at Tevatron Compatible with a Hadron Molecule Interpretation?}",
    archivePrefix = "arXiv",
    primaryClass = "hep-ph",
    doi = "10.1103/PhysRevLett.103.162001",
    journal = "Phys. Rev. Lett.",
    volume = "103",
    pages = "162001",
    year = "2009"
}

@article{Fleming:2011xa,
    author = "Fleming, Sean and Mehen, Thomas",
    title = "{The decay of the $X(3872)$ into $\chi_{cJ}$ and the Operator Product Expansion in XEFT}",
    archivePrefix = "arXiv",
    primaryClass = "hep-ph",
    reportNumber = "INT-PUB-11-042",
    doi = "10.1103/PhysRevD.85.014016",
    journal = "Phys. Rev. D",
    volume = "85",
    pages = "014016",
    year = "2012"
}

@article{Aceti:2012cb,
    author = "Aceti, F. and Molina, R. and Oset, E.",
    title = "{The $X(3872) \to J/\psi \gamma$ decay in the $D \bar D^*$ molecular picture}",
    archivePrefix = "arXiv",
    primaryClass = "hep-ph",
    doi = "10.1103/PhysRevD.86.113007",
    journal = "Phys. Rev. D",
    volume = "86",
    pages = "113007",
    year = "2012"
}

@article{Margaryan:2013tta,
    author = "Margaryan, Arman and Springer, Roxanne P.",
    title = "{Using the decay {\ensuremath{\psi}}(4160){\textrightarrow}X(3872){\ensuremath{\gamma}} to probe the molecular content of the X(3872)}",
    archivePrefix = "arXiv",
    primaryClass = "hep-ph",
    doi = "10.1103/PhysRevD.88.014017",
    journal = "Phys. Rev. D",
    volume = "88",
    number = "1",
    pages = "014017",
    year = "2013"
}

@article{Guo:2013zbw,
    author = "Guo, Feng-Kun and Hanhart, Christoph and Mei{\ss}ner, Ulf-G. and Wang, Qian and Zhao, Qiang",
    title = "{Production of the X(3872) in charmonia radiative decays}",
    archivePrefix = "arXiv",
    primaryClass = "hep-ph",
    doi = "10.1016/j.physletb.2013.06.053",
    journal = "Phys. Lett. B",
    volume = "725",
    pages = "127--133",
    year = "2013"
}

@article{Guo:2014hqa,
    author = "Guo, F. K. and Hidalgo-Duque, C. and Nieves, J. and Ozpineci, Altug and Valderrama, M. P.",
    title = "{Detecting the long-distance structure of the $X$(3872)}",
    archivePrefix = "arXiv",
    primaryClass = "hep-ph",
    doi = "10.1140/epjc/s10052-014-2885-4",
    journal = "Eur. Phys. J. C",
    volume = "74",
    number = "5",
    pages = "2885",
    year = "2014"
}

@article{Zhou:2019swr,
    author = "Zhou, Zhi-Yong and Yu, Meng-Ting and Xiao, Zhiguang",
    title = "{Decays of $X(3872)$ to $\chi_{cJ}\pi^0$ and $J/\psi\pi^+\pi^-$}",
    archivePrefix = "arXiv",
    primaryClass = "hep-ph",
    reportNumber = "USTC-ICTS-19-08",
    doi = "10.1103/PhysRevD.100.094025",
    journal = "Phys. Rev. D",
    volume = "100",
    number = "9",
    pages = "094025",
    year = "2019"
}

@article{Wu:2021udi,
    author = "Wu, Qi and Chen, Dian-Yong and Matsuki, Takayuki",
    title = "{A phenomenological analysis on isospin-violating decay of $X(3872)$}",
    archivePrefix = "arXiv",
    primaryClass = "hep-ph",
    doi = "10.1140/epjc/s10052-021-08984-2",
    journal = "Eur. Phys. J. C",
    volume = "81",
    number = "2",
    pages = "193",
    year = "2021"
}

@article{Meng:2021kmi,
    author = "Meng, Lu and Wang, Guang-Juan and Wang, Bo and Zhu, Shi-Lin",
    title = "{Revisit the isospin violating decays of X(3872)}",
    archivePrefix = "arXiv",
    primaryClass = "hep-ph",
    doi = "10.1103/PhysRevD.104.094003",
    journal = "Phys. Rev. D",
    volume = "104",
    number = "9",
    pages = "094003",
    year = "2021"
}

@article{Wang:2022qxe,
    author = "Wang, Yan and Wu, Qi and Li, Gang and Qin, Wen-Hua and Liu, Xiao-Hai and An, Chun-Sheng and Xie, Ju-Jun",
    title = "{Investigations of charmless decays of X(3872) via intermediate meson loops}",
    archivePrefix = "arXiv",
    primaryClass = "hep-ph",
    doi = "10.1103/PhysRevD.106.074015",
    journal = "Phys. Rev. D",
    volume = "106",
    number = "7",
    pages = "074015",
    year = "2022"
}

@article{Wu:2023rrp,
    author = "Wu, Qi and Liu, Ming-Zhu and Geng, Li-Sheng",
    title = "{Productions of X(3872), $Z_c(3900)$, $X_2(4013)$, and $Z_c(4020)$ in $B_{(s)}$ decays offer strong clues on their molecular nature}",
    archivePrefix = "arXiv",
    primaryClass = "hep-ph",
    doi = "10.1140/epjc/s10052-024-12501-6",
    journal = "Eur. Phys. J. C",
    volume = "84",
    number = "2",
    pages = "147",
    year = "2024"
}

@inproceedings{Belle:2005lfc,
    author = "Abe, Kazuo and others",
    collaboration = "Belle Collaboration",
    title = "{Evidence for X(3872) ---{\ensuremath{>}} gamma J / psi and the sub-threshold decay X(3872) ---{\ensuremath{>}} omega J / psi}",
    booktitle = "{Procceedings of the 22nd International Symposium on Lepton-Photon Interactions at High Energy (LP 2005)}",
    eprint = "arXiv:hep-ex/0505037",
    archivePrefix = "arXiv",
    reportNumber = "BELLE-CONF-0540, LP-2005-175",
    month = "5",
    year = "2005"
}

@article{BESIII:2019qvy,
    author = "Ablikim, M. and others",
    collaboration = "BESIII Collaboration",
    title = "{Study of $e^+e^- \to \gamma \omega J/\psi$ and Observation of $X(3872) \to \omega J/\psi$}",
    eprint = "1903.04695",
    archivePrefix = "arXiv",
    primaryClass = "hep-ex",
    doi = "10.1103/PhysRevLett.122.232002",
    journal = "Phys. Rev. Lett.",
    volume = "122",
    number = "23",
    pages = "232002",
    year = "2019"
}

@article{BaBar:2010wfc,
    author = "del Amo Sanchez, P. and others",
    collaboration = "\textit{BaBar} Collaboration",
    title = "{Evidence for the decay X(3872) ---{\ensuremath{>}} J/ psi omega}",
    archivePrefix = "arXiv",
    primaryClass = "hep-ex",
    reportNumber = "SLAC-PUB-14143, BABAR-PUB-10-009",
    doi = "10.1103/PhysRevD.82.011101",
    journal = "Phys. Rev. D",
    volume = "82",
    pages = "011101",
    year = "2010"
}

@article{LHCb:2022jez,
    author = "Aaij, Roel and others",
    collaboration = "LHCb Collaboration",
    title = "{Observation of sizeable {\ensuremath{\omega}} contribution to {\ensuremath{\chi}}c1(3872){\textrightarrow}{\ensuremath{\pi}}+{\ensuremath{\pi}}-J/{\ensuremath{\psi}} decays}",
    archivePrefix = "arXiv",
    primaryClass = "hep-ex",
    reportNumber = "LHCb-PAPER-2021-045, CERN-EP-2022-049",
    doi = "10.1103/PhysRevD.108.L011103",
    journal = "Phys. Rev. D",
    volume = "108",
    number = "1",
    pages = "L011103",
    year = "2023"
}

@article{Yamaguchi:2019vea,
    author = "Yamaguchi, Yasuhiro and Hosaka, Atsushi and Takeuchi, Sachiko and Takizawa, Makoto",
    title = "{Heavy hadronic molecules with pion exchange and quark core couplings: a guide for practitioners}",
    archivePrefix = "arXiv",
    primaryClass = "hep-ph",
    reportNumber = "RIKEN-QHP-425",
    doi = "10.1088/1361-6471/ab72b0",
    journal = "J. Phys. G",
    volume = "47",
    number = "5",
    pages = "053001",
    year = "2020"
}

@article{Wang:2023ovj,
    author = "Wang, Guang-Juan and Yang, Zhi and Wu, Jia-Jun and Oka, Makoto and Zhu, Shi-Lin",
    title = "{New insight into the exotic states strongly coupled with the DD£þ{\ensuremath{*}} from the Tcc+}",
    archivePrefix = "arXiv",
    primaryClass = "hep-ph",
    doi = "10.1016/j.scib.2024.07.012",
    journal = "Sci. Bull.",
    volume = "69",
    pages = "3036--3041",
    year = "2024"
}

@article{Song:2023pdq,
    author = "Song, Jing and Dai, L. R. and Oset, E.",
    title = "{Evolution of compact states to molecular ones with coupled channels: The case of the X(3872)}",
    archivePrefix = "arXiv",
    primaryClass = "hep-ph",
    doi = "10.1103/PhysRevD.108.114017",
    journal = "Phys. Rev. D",
    volume = "108",
    number = "11",
    pages = "114017",
    year = "2023"
}

@article{Gamermann:2009uq,
    author = "Gamermann, D. and Nieves, J. and Oset, E. and Ruiz Arriola, E.",
    title = "{Couplings in coupled channels versus wave functions: application to the X(3872) resonance}",
    archivePrefix = "arXiv",
    primaryClass = "hep-ph",
    doi = "10.1103/PhysRevD.81.014029",
    journal = "Phys. Rev. D",
    volume = "81",
    pages = "014029",
    year = "2010"
}

@article{Gamermann:2007fi,
    author = "Gamermann, D. and Oset, E.",
    title = "{Axial resonances in the open and hidden charm sectors}",
    archivePrefix = "arXiv",
    primaryClass = "hep-ph",
    doi = "10.1140/epja/i2007-10435-1",
    journal = "Eur. Phys. J. A",
    volume = "33",
    pages = "119--131",
    year = "2007"
}

@article{Gamermann:2007mu,
    author = "Gamermann, D. and Oset, E.",
    title = "{Hidden charm dynamically generated resonances and the e+ e- ---{\ensuremath{>}} J / psi D anti-D, J / psi D anti-D* reactions}",
    archivePrefix = "arXiv",
    primaryClass = "hep-ph",
    doi = "10.1140/epja/i2007-10580-5",
    journal = "Eur. Phys. J. A",
    volume = "36",
    pages = "189--194",
    year = "2008"
}

@article{Gamermann:2009fv,
    author = "Gamermann, Daniel and Oset, Eulogio",
    title = "{Isospin breaking effects in the X(3872) resonance}",
    archivePrefix = "arXiv",
    primaryClass = "hep-ph",
    doi = "10.1103/PhysRevD.80.014003",
    journal = "Phys. Rev. D",
    volume = "80",
    pages = "014003",
    year = "2009"
}

@article{Gamermann:2010ga,
    author = "Gamermann, D. and Jimenez-Tejero, C. E. and Ramos, A.",
    title = "{Radiative decays of dynamically generated charmed baryons}",
    archivePrefix = "arXiv",
    primaryClass = "hep-ph",
    doi = "10.1103/PhysRevD.83.074018",
    journal = "Phys. Rev. D",
    volume = "83",
    pages = "074018",
    year = "2011"
}

@article{Dong:2009yp,
    author = "Dong, Yubing and Faessler, Amand and Gutsche, Thomas and Kovalenko, Sergey and Lyubovitskij, Valery E.",
    title = "{X(3872) as a hadronic molecule and its decays to charmonium states and pions}",
    archivePrefix = "arXiv",
    primaryClass = "hep-ph",
    doi = "10.1103/PhysRevD.79.094013",
    journal = "Phys. Rev. D",
    volume = "79",
    pages = "094013",
    year = "2009"
}

@article{BESIII:2019esk,
    author = "Ablikim, M. and others",
    collaboration = "BESIII Collaboration",
    title = "{Observation of the decay $X(3872) \to \pi^0 \chi_{c1}(1P)$}",
    archivePrefix = "arXiv",
    primaryClass = "hep-ex",
    doi = "10.1103/PhysRevLett.122.202001",
    journal = "Phys. Rev. Lett.",
    volume = "122",
    number = "20",
    pages = "202001",
    year = "2019"
}

@article{ParticleDataGroup:2024cfk,
    author = "Navas, S. and others",
    collaboration = "Particle Data Group",
    title = "{Review of particle physics}",
    doi = "10.1103/PhysRevD.110.030001",
    journal = "Phys. Rev. D",
    volume = "110",
    number = "3",
    pages = "030001",
    year = "2024"
}

@article{BESIII:2022kow,
    author = "Ablikim, M. and others",
    collaboration = "BESIII Collaboration",
    title = "{Search for$X(3872)\to\pi^0\chi_{c0}$ and $X(3872)\to\pi\pi\chi_{c0}$ at BESIII}",
    archivePrefix = "arXiv",
    primaryClass = "hep-ex",
    doi = "10.1103/PhysRevD.105.072009",
    journal = "Phys. Rev. D",
    volume = "105",
    number = "7",
    pages = "072009",
    year = "2022"
}

@article{BESIII:2023eeb,
    author = "Ablikim, M. and others",
    collaboration = "BESIII Collaboration",
    title = "{Search for the decay $\chi_{c1}(3872)\to\pi^{+}\pi^{-}\chi_{c1}$}",
    archivePrefix = "arXiv",
    primaryClass = "hep-ex",
    doi = "10.1103/PhysRevD.109.L071101",
    journal = "Phys. Rev. D",
    volume = "109",
    number = "7",
    pages = "L071101",
    year = "2024"
}

@article{BESIII:2024ilt,
    author = "Ablikim, Medina and others",
    collaboration = "BESIII Collaboration",
    title = "{Search for X(3872){\textrightarrow}{\ensuremath{\pi}}0{\ensuremath{\pi}}0{\ensuremath{\chi}}c1,2}",
    archivePrefix = "arXiv",
    primaryClass = "hep-ex",
    doi = "10.1103/PhysRevD.110.072015",
    journal = "Phys. Rev. D",
    volume = "110",
    number = "7",
    pages = "072015",
    year = "2024"
}

@article{Achasov:2024anu,
    author = "Achasov, N. N. and Shestakov, G. N.",
    title = "{Tentative estimates of B(X(3872){\textrightarrow}{\ensuremath{\pi}}0{\ensuremath{\pi}}0{\ensuremath{\chi}}c1) and B(X(3872){\textrightarrow}{\ensuremath{\pi}}+{\ensuremath{\pi}}-{\ensuremath{\chi}}c1)}",
    archivePrefix = "arXiv",
    primaryClass = "hep-ph",
    doi = "10.1103/PhysRevD.110.016023",
    journal = "Phys. Rev. D",
    volume = "110",
    number = "1",
    pages = "016023",
    year = "2024"
}

@article{Jia:2023pud,
    author = "Jia, Zhao-Sai and Zhang, Zhen-Hua and Qin, Wen-Hua and Li, Gang",
    title = "{Hunting for Xb via hidden bottomonium decays Xb{\textrightarrow}{\ensuremath{\pi}}{\ensuremath{\pi}}{\ensuremath{\chi}}bJ}",
    archivePrefix = "arXiv",
    primaryClass = "hep-ph",
    doi = "10.1103/PhysRevD.109.034017",
    journal = "Phys. Rev. D",
    volume = "109",
    number = "3",
    pages = "034017",
    year = "2024"
}

@article{Liu:2024ogo,
    author = "Liu, Shi-Dong and Wang, Fan and Jia, Zhao-Sai and Li, Gang and Liu, Xiao-Hai and Xie, Ju-Jun",
    title = "{Pionic transitions of the spin-2 partner of X(3872) to {\ensuremath{\chi}}cJ}",
    archivePrefix = "arXiv",
    primaryClass = "hep-ph",
    doi = "10.1103/PhysRevD.110.054048",
    journal = "Phys. Rev. D",
    volume = "110",
    number = "5",
    pages = "054048",
    year = "2024"
}

@article{Cai:2025inq,
    author = "Cai, Hao-Dong and Jia, Zhao-Sai and Li, Gang and Liu, Shi-Dong",
    title = "{Hidden charmed decays of X(3872) within the DD{\textasciimacron}* molecular framework}",
    archivePrefix = "arXiv",
    primaryClass = "hep-ph",
    doi = "10.1103/gn25-fc9q",
    journal = "Phys. Rev. D",
    volume = "111",
    number = "11",
    pages = "114024",
    year = "2025"
}

@article{Guo:2009wr,
    author = "Guo, Feng-Kun and Hanhart, Christoph and Meissner, Ulf-G.",
    title = "{On the extraction of the light quark mass ratio from the decays psi-prime ---{\ensuremath{>}} J/psi pi0 (eta)}",
    archivePrefix = "arXiv",
    primaryClass = "hep-ph",
    reportNumber = "FZJ-IKP-TH-2009-22, HISKP-TH-09-23",
    doi = "10.1103/PhysRevLett.103.082003",
    journal = "Phys. Rev. Lett.",
    volume = "103",
    pages = "082003",
    year = "2009",
    note = "\textbf{104}, 109901 (2010)"
}

@article{Guo:2010ak,
    author = "Guo, Feng-Kun and Hanhart, Christoph and Li, Gang and Meissner, Ulf-G. and Zhao, Qiang",
    title = "{Effect of charmed meson loops on charmonium transitions}",
    archivePrefix = "arXiv",
    primaryClass = "hep-ph",
    reportNumber = "FZJ-IKP-TH-2010-08, HISKP-TH-10-09",
    doi = "10.1103/PhysRevD.83.034013",
    journal = "Phys. Rev. D",
    volume = "83",
    pages = "034013",
    year = "2011"
}

@article{Wise:1992hn,
    author = "Wise, Mark B.",
    title = "{Chiral perturbation theory for hadrons containing a heavy quark}",
    reportNumber = "CALT-68-1765",
    doi = "10.1103/PhysRevD.45.R2188",
    journal = "Phys. Rev. D",
    volume = "45",
    number = "7",
    pages = "R2188",
    year = "1992"
}

@article{Hu:2005gf,
    author = "Hu, Jie and Mehen, Thomas",
    title = "{Chiral Lagrangian with heavy quark-diquark symmetry}",
    archivePrefix = "arXiv",
    reportNumber = "JLAB-THY-05-452",
    doi = "10.1103/PhysRevD.73.054003",
    journal = "Phys. Rev. D",
    volume = "73",
    pages = "054003",
    year = "2006"
}

@article{Mehen:2015efa,
    author = "Mehen, Thomas",
    title = "{Hadronic loops versus factorization in effective field theory calculations of X(3872) {\textrightarrow} $¦Ö_{cJ}¦Ð^0$}",
    archivePrefix = "arXiv",
    primaryClass = "hep-ph",
    doi = "10.1103/PhysRevD.92.034019",
    journal = "Phys. Rev. D",
    volume = "92",
    number = "3",
    pages = "034019",
    year = "2015"
}

@article{Weinberg:1965zz,
    author = "Weinberg, Steven",
    title = "{Evidence That the Deuteron Is Not an Elementary Particle}",
    doi = "10.1103/PhysRev.137.B672",
    journal = "Phys. Rev.",
    volume = "137",
    pages = "B672--B678",
    year = "1965"
}

@article{Baru:2003qq,
    author = "Baru, V. and Haidenbauer, J. and Hanhart, C. and Kalashnikova, Yu. and Kudryavtsev, Alexander Evgenyevich",
    title = "{Evidence that the a(0)(980) and f(0)(980) are not elementary particles}",
    archivePrefix = "arXiv",
    reportNumber = "FZJ-IKP-TH-2003-13",
    doi = "10.1016/j.physletb.2004.01.088",
    journal = "Phys. Lett. B",
    volume = "586",
    pages = "53--61",
    year = "2004"
}

@article{Colangelo:2003sa,
    author = "Colangelo, P. and De Fazio, F. and Pham, T. N.",
    title = "{Nonfactorizable contributions in B decays to charmonium: The Case of B- ---{\ensuremath{>}} K- h(c)}",
    archivePrefix = "arXiv",
    reportNumber = "BARI-TH-03-467",
    doi = "10.1103/PhysRevD.69.054023",
    journal = "Phys. Rev. D",
    volume = "69",
    pages = "054023",
    year = "2004"
}

@article{Jia:2025mtn,
    author = "Jia, Zhao-Sai and Li, Gang and Zhang, Zhen-Hua",
    title = "{Constrain the $¦Ö_{cJ}\to D^{(*)}\bar{D}^{(*)}$ effective couplings via the $X(3872)\to ¦Ð^0¦Ö_{cJ}$ decays}",
    archivePrefix = "arXiv",
    journal = {Phys. Rev. D},
  volume = {112},
  issue = {7},
  pages = {074035},
  numpages = {8},
  year = {2025},
  month = {Oct},
  publisher = {American Physical Society},
  doi = {10.1103/51z1-qls5},
}

@article{Guo:2016yxl,
author = "Guo, Feng-Kun and Mei{\ss}ner, Ulf-G. and Yang, Zhi",
title = "{Hindered magnetic dipole transitions between P-wave bottomonia and coupled-channel effects}",
archivePrefix = "arXiv",
primaryClass = "hep-ph",
doi = "10.1016/j.physletb.2016.07.023",
journal = "Phys. Lett. B",
volume = "760",
pages = "417--421",
year = "2016"
}

@article{Chen:2019gfp,
    author = "Chen, Yun-Hua",
    title = "{Predictions of $\Upsilon(4S) \to h_b(1P,2P) \pi^+\pi^-$ transitions}",
    archivePrefix = "arXiv",
    primaryClass = "hep-ph",
    doi = "10.1088/1674-1137/44/2/023103",
    journal = "Chin. Phys. C",
    volume = "44",
    number = "2",
    pages = "023103",
    year = "2020"
}

@article{Tang:2023yls,
    author = "Tang, Meng-Na and Lin, Yong-Hui and Guo, Feng-Kun and Hanhart, Christoph and Mei{\ss}ner, Ulf-G.",
    title = "{Isospin-conserving hadronic decay of the D $_{s1}$(2460) into D $_{s}$ {\ensuremath{\pi}} $^{+}$ {\ensuremath{\pi}} $^{−}$}",
    archivePrefix = "arXiv",
    primaryClass = "hep-ph",
    doi = "10.1088/1572-9494/accc1f",
    journal = "Commun. Theor. Phys.",
    volume = "75",
    number = "5",
    pages = "055203",
    year = "2023"
}
%\nocite{*}

\end{CJK}	
\end{document}